\documentclass[a4paper,11pt]{article}
\pdfoutput=1 
\usepackage{jheppub}
 \usepackage{mathtools, booktabs, caption, float, multirow, comment} 
\usepackage{multirow}

\newtheorem{defn}{Definition}[section]
\usepackage{longtable}
\usepackage{enumitem}
\usepackage{array}
\makeatletter
\gdef\@fpheader{\text{ }}
\makeatother

\newcommand{\ba}{\begin{align*}}
\newcommand{\ea}{\end{align*}}
\newcommand{\ceff}{c_{\text{eff}}}
\newcommand{\heff}{H_{\text{eff}}}
\newcommand{\BZ}{\mathbb{Z}}
\newcommand{\psl}[1]{\text{PSL}(2,#1)}

\title{\boldmath Updating the holomorphic modular bootstrap}
\author{Suresh Govindarajan and Akhila Sadanandan}
  
\affiliation{Department of Physics,
 Indian Institute of Technology Madras,
 Chennai 600036, India\\ and \\
Centre for Operator Algebras, Geometry, Matter and Spacetime, \\Indian Institute of Technology Madras, Chennai 600036 India}

\preprint{v1.1 April, 2026}
\arxivnumber{2604.11277}

\emailAdd{suresh@physics.iitm.ac.in}
\emailAdd{akhila@physics.iitm.ac.in }

\abstract{We update the holomorphic modular bootstrap incorporating a 
recent result that computes the exact $S$-matrix within the Modular 
Linear Differential Equation (MLDE) setting. Further, using knowledge of 
the allowed exponents modulo one, we obtain admissible solutions to all 
MLDE's with up to six characters and Wronskian index $<6$ and one 
accessory parameter with $c_\text{eff}\leq 24$. We then identify which 
of the admissible solutions have good fusion rules -- we call such 
solutions \textit{tenable}. When possible, we identify the CFT and in 
the unitary cases the MTC class they belong to.}

\begin{document}
\setcounter{tocdepth}{2}
\maketitle
\flushbottom

 \definecolor{mgreen}{rgb}{0,0.5,0}

\clearpage

\textit{If someone hands you a $q$-character vector, and says it looks
like one from a rational VOA, and passes all the obvious tests,
don’t believe them!! -- Terry Gannon\cite{Gannon:2025}}

\section{Introduction}

 Two-dimensional Conformal Field Theories (CFTs) play a central role in 
 a wide range of physical and mathematical contexts, ranging from 
 quantum hall systems to the AdS/CFT correspondence. Rational Conformal 
 Field Theories(RCFTs) are CFTs with a finite number of primaries 
 \cite{Anderson:1987ge}. The partition function of an RCFT on a torus 
 takes the following form
\begin{equation}
    Z(\tau,\bar{\tau}) = \sum_{i,j=0}^{n-1} M_{ij}\ \bar\chi_i(\bar\tau)\ \chi_j(\tau)\ ,
\end{equation}
where $\chi_i(\tau)$ are the characters associated with the Verma module 
$V_i$ with $i=0$ representing the one associated with the identity 
operator. Let $c$ denote the central charge and $(h_i,\bar{h}_i)$ the 
conformal weights of the RCFT. The characters have following $q$-series
\begin{equation}
    \chi_j(\tau) := \text{Tr}_{V_j}\left(q^{L_0-\frac{c}{24}}\right) = q^{\alpha_j}\ \sum_{m=0}^\infty a_{j,m}\ q^m\ ,
\end{equation}
where $a_{0,0}=1$ and all $a_{i,m}$ are non-negative integers and the 
exponents $\alpha_i = (h_i-\frac{c}{24})$.
 
The characters of an RCFT form a Vector Valued Modular Form (VVMF) of 
weight zero. Let $\mathbb{X}=(\chi_0,\ldots,\chi_{n-1})^T$ denote the 
VVMF. The modular transformation of the VVMF is given through a 
multiplier $v:\psl{\BZ}\rightarrow \text{GL}(n,\mathbb{C})$. One has
\[
\mathbb{X}\left(\frac{a\tau+b}{c\tau+d}\right) = v\left(\begin{smallmatrix}
a & b \\ c & d 
\end{smallmatrix}\right)\cdot \mathbb{X}(\tau)\quad,\ 
\left(\begin{smallmatrix}
a & b \\ c & d 
\end{smallmatrix}\right) \in \psl{\BZ}\ .
 \]
Let $T$ denote the matrix $v\left(\begin{smallmatrix} 1 & 1 \\ 1 & 0 
\end{smallmatrix}\right)$ and $S$ denote the matrix 
$v\left(\begin{smallmatrix} 0 & -1 \\ 1 & 0 \end{smallmatrix}\right)$. 
These two matrices determine the multiplier\cite{Gannon:2013jua}. It is 
easy to see that the $T$ matrix is determined by the exponents. One has 
$T=\text{Diag}(e^{2\pi i \alpha_0},\ldots,e^{2\pi i \alpha_{n-1}})$. 
Modular invariance of the partition of the torus implies
\[
T^\dagger \cdot M \cdot T = M\quad,\quad S^\dagger \cdot M \cdot S = M\ .
\]
We will focus mostly on cases where $M$ is a diagonal matrix with 
entries $(m_0=1,m_1,\ldots,m_{n-1})$.

The modular property can be used to interpret the characters as 
solutions to the Modular Linear Differential Equation(MLDE).
 \begin{equation}\label{MLDEgen}
 \Big(D^{(n)}  + \sum_{s=1}^n \phi_{2s}(\tau)\ D^{(n-s)}\Big)\ \chi_i(\tau)=0\ ,
 \end{equation}
 where $D$ is the Ramanujan-Serre covariant derivative and 
 $\phi_{2s}(\tau)$ are weakly holomorphic modular forms of weight $2s$. 
 This forms a basis for the Mathur-Mukhi-Sen (MMS) proposal for the 
 classification of RCFTs that is called the holomorphic modular 
 bootstrap\cite{Mathur:1988gt,Mathur:1988na,Naculich:1988xv}. Two 
 important parameters in the classification of RCFTs are the number of 
 characters ($n$), and the Wronskian index ($\ell$). The Wronskian index 
 is related to the exponents via the formula
\begin{equation}
    \ell = \frac{n(n-1)}{2}-6 \sum_{i=0}^{n-1} \alpha_i \ ,
\end{equation}
and it determines the pole structure of the MLDE. Fixing $(n,\ell)$ 
determines the MLDE up to a finite number of parameters. The holomorphic 
modular bootstrap aims to determine the parameters that lead to 
solutions that have a $q$-series with non-negative integral 
coefficients.
\begin{defn}
    A solution to an MLDE is called admissible if the $q$-series for all 
    its solutions are non-negative integers. Further, the leading 
    coefficient of the identity character must be unity.
\end{defn}
 
In recent years, the MMS program has experienced a significant revival, 
specifically in the classification of theories for a small number of 
characters and vanishing Wronskian index. For two-character theories 
($n=2$), the $\ell=0$ case was originally classified by MMS, while 
extensions to higher Wronskian indices such as $\ell=2, 4$ and beyond 
was first studied in \cite{Naculich:1988xv} and then extensively 
explored by Mukhi and collaborators 
\cite{Hampapura:2015cea,Chandra:2018pjq,Mukhi:2022bte}. Three-character 
theories ($n=3$), with vanishing Wronskian index are studied in 
\cite{Franc:2016,Franc:2020,Das:2021uvd,Kaidi:2021ent,Bae:2021mej}. and 
examples with non-zero Wronskian index are constructed from MLDE 
approach in \cite{Gowdigere:2023xnm}.

Looking for admissible solutions to an MLDE is a Diophantine problem 
that becomes harder as the number of parameters in the MLDE increases. 
This increase can happen either due to an increase in the number of 
characters and/or the Wronskian index. Kaidi, Lin, and Parra-Martinez 
(KLP) provided a simplification by identifying the allowed exponents 
modulo one using modular representation theory -- this is equivalent to 
giving the $T$-matrix associated with the solution\cite{Kaidi:2021ent}. 
For rigid MLDEs, i.e., those without accessory parameters, one can 
reduce the search space to a finite (but possibly large) set obtained by 
fixing the modulo one ambiguity with a constraint on the central charge. 
Then we can look for admissible solutions within this finite set. KLP 
list allowed exponents modulo one for $n\leq 5$. In such cases, the 
MLDEs with a vanishing Wronskian index are rigid. In such cases, KLP 
provide a list of admissible solutions with suitable constraints imposed 
on central charge. Rayhaun combined data from Modular Tensor 
Categories\cite{Ng:2023} as well as the theory of vector valued modular 
forms\cite{Bantay:2007zz, Gannon:2013jua}, Rayhaun solved for admissible 
solutions with $c\leq24$ and up to four primaries\cite{Rayhaun:2023pgc}.

Given an admissible solution, one would like to know its modular 
properties. In particular, we need to determine the $S$-matrix of the 
admissible solution. Then we can use the Verlinde formula to obtain the 
fusion rules. One can then check if the fusion rules are non-negative 
and compatible with an RCFT.
\begin{defn}
An admissible solution is called \textit{tenable} if the fusion rules 
obtained from the Verlinde formula are compatible with those of an RCFT.
\end{defn}
It is important to emphasize that every tenable solution need not to 
correspond to an RCFT. Gannon has provided examples of tenable solutions 
with central charges $c=8,16$ that are candidates for a Haagerup 
RCFT\cite{Gannon:2025}. He shows that these cannot be RCFTs by making 
use of the extended Kac-Moody symmetry in these examples.
 
In some cases, the MLDE is of hypergeometric type and the $S$-matrix is 
immediately available through the work of Beukers and 
Heckman\cite{Beukers1989}. For instance, MLDEs with $n\leq 3$ and 
vanishing Wronskian index fall into this class. Using a group theoretic 
approach due to Mathur-Sen\cite{Mathur:1989pk}, KLP work out the case of 
$(4,0)$ MLDEs where they are able to determine the possible monodromy 
groups and thereby the $S$-matrices. This approach gets harder as we go 
to higher number of characters. A recent result enables us to obtain the 
$S$-matrix directly from the MLDE in a fairly straightforward 
fashion\cite{Govindarajan:2026frs}. This is an important simplification 
and we update the holomorphic modular bootstrap by incorporating this 
result. The updated holomorphic modular bootstrap now consists of the 
following steps.
\begin{itemize}

\item[Step 1:] Fix the number of characters $n$ and the Wronskian index 
$\ell<6$\footnote{An important feature of MLDE's with Wronskian index 
$\ell<6$, is that the singularities occur at $q=0,i,e^{2\pi i/3}$. When 
$\ell\geq 6$, singularities can occur in the interior of the moduli 
space. This is important for Step 3.}. Use the KLP method to obtain 
exponents modulo one. Fix the modulo one ambiguity by imposing a 
constraint on the central charge.
\item[Step 2:] When there are accessory parameters, solve the 
Diophantine equations that simultaneously fix the accessory parameters 
and lead to admissible solutions. There is an ambiguity associated with 
the degeneracy of admissible solutions that we fix in the next step.
\item[Step 3:] Compute the $S$-matrix as well as the multiplicity of 
each character. One way to fix the ambiguity mentioned in Step 2 is to 
require that the multiplicities are square-free positive integers.  We 
then determine which of the admissible solutions are tenable.
\end{itemize}
We illustrate the method for MLDEs with one accessory parameter. These 
are the $(4,2)$, $(5,2)$, and $(6,0)$ MLDEs. An interesting case is that 
of the $(4,4)$ MLDE which is expected to have two accessory parameters, 
but we find that one of them is always zero. We solve for such examples 
as well. The main result of this paper is a complete list of admissible 
solutions with effective central charge $\ceff\leq 24$ in all these 
cases. We then identify which among the admissible solutions are 
tenable.

The organization of the paper is as follows. Section 1 introduces the 
problem as well as states the main ideas of the paper. In section 2, we 
review the MLDE approach to the holomorphic modular bootstrap and 
discuss the MLDEs that we study in this paper. In section 3, we discuss 
the two major updates to the bootstrap i.e., the KLP method of 
determining the exponents and the determination of the $S$-matrix and 
the ambiguities associated with it. This is explicitly illustrated by a 
specific $(4,2)$ example where all details are given. In section 4, we 
review other approaches to resolving the ambiguities mentioned in 
section 3. We conclude in section 4 with an extended discussion and some 
concluding remarks. Appendix A discusses the notation and modular 
objects that we use. Appendix B gives the MLDEs of interest in the 
$w$-plane. Appendix C gives the list of exponents modulo 1 for the six 
character case. Appendix D contains the explicit results of our work. 
Various tables in this appendix list the admissible solutions clearly 
indicating which ones are tenable, which ones have multiple vacua and 
which ones are not tenable. We also provide a list of $S$-matrices for 
all cases except for untenable ones. We also identify when a solution is 
potentially a unitary RCFT and when a solution is potentially 
non-unitary by indicating which index represents the primary.

\section{Modular Linear Differential Equations}

\subsection{Determining the MLDEs of interest}

Let us review how one determines the parameters associated with a MLDE 
of type $(n,\ell)$ -- we follow the discussion in \cite{Das:2023qns}. 
Recall that the Eisenstein series, $(E_4(\tau),E_6(\tau))$, provide a 
basis for the ring of holomorphic modular forms of $PSL(2,\mathbb{Z})$ 
while the ring of weakly holomorphic modular forms is generated by 
$\mathbb{C}(J,E_4,E_6)$\cite{Gannon:2013jua}. Thus, the 
$\phi_{2s}(\tau)$ which are weakly holomorphic modular forms of weight 
$2s$ can be constructed using this result. The Wronskian index equals 
six times the number zeros of the Wronskian of the solutions of the 
MLDE. The zeros at the elliptic points $\tau=i, \exp(2\pi i/3)$ are 
divided by the order of the orbifold singularity at these points. Thus a 
zero at $\tau=i$ will contribute $3$ to the Wronskian index and a zero 
at $\tau=\exp(2\pi i/3)$ will contribute $2$ to the Wronskian index. 
Thus, when $\ell=0$, all the $\phi_{2s}$ are necessarily holomorphic. It 
is not possible to have the Wronskian index to equal one. When 
$\ell=2,4$, one can allow a non-holomorphicity via single power of 
$E_4(\tau)$ (which has a zero at $\tau=\exp(2\pi i/3)$) in the 
denominator of $\phi_{2s}$. Similarly, for $\ell=3$ one call allow a 
non-holmorphicity via single power of $E_6(\tau)$ (which has a zero at 
$\tau=i$) in the denominator of $\phi_{2s}$. Finally, when $\ell=5$, one 
can allow for denominators involving a single power of $E_4$ and $E_6$. 
This enables us to fix the form of the MLDEs that we study in this paper. 
Additional conditions can arise from the monodromy about the elliptic 
points.

The MLDE associated with characters of type $(n,\ell)$ is said to be 
rigid if it is completely determined by its exponents. MLDEs associated 
with vanishing Wronskian index are rigid for all $n\leq 5$.  MLDEs that 
are not rigid have additional parameters that are not determined by the 
exponents. They are called \textit{accessory} parameters. We also 
restrict our considerations to the Wronskian index $\ell<6$.  We study 
the following non-rigid MLDEs here: $(4,2)$ , $(4,4)$, $(5,2)$ and 
$(6,0)$ as these examples have one accessory parameter.  The MLDEs of 
type $(3,3)$ were the first one with one accessory parameter for which 
all admissible solutions with $c\leq96$ were classified. This involves 
solving a Diophantine equation for four parameters which required some 
ingenuity to find the solutions of interest.

The modular linear differential operators that we study are given in 
Table \ref{MLDO}
\begin{table}[h]
\begin{tabular}{c|l}
\centering
$(n,\ell)$ & The modular linear differential operator (MLDO) \\ \hline
 $(4,2)$    & $\left[ D^{(4)}+ \frac13 \frac{E_6}{E_4} D^{(3)} + \kappa E_4 D^{(2)}  + \mu E_6 D^{(1)} +(\nu E_4^2+\rho \frac{\Delta}{E_4})\right] $   \\ \midrule
 $(4,4)$    & $\left[ D^{(4)} + \frac23\frac{E_6}{E_4}D^{(3)}  +\left (\kappa E_4+\rho_1\frac{\Delta}{E_4^2}\right) D^{(2)}+ \mu E_6 D^{(1)} +\left(\nu E_4^2+\rho \frac{\Delta}{E_4}\right)\right]$ \\ \midrule
 $(5,2)$ & $\left[ D^{(5)} + \frac13\frac{E_6}{E_4}D^{(4)}+\mu_2 E_4 D^{(3)}+ \mu_3 E_6 D^{(2)}  + (\mu_4 E_4^2 +\rho \frac{\Delta}{E_4})D^{(1)} +\nu E_4E_6\right]$ \\ \midrule
 $(6,0)$ & $\left[ D^{(6)} + \kappa E_4 D^{(4)}+ \mu_3 E_6 D^{(3)}+ \mu_4 E_4^2 D^{(2)}  + \mu_5 E_4E_6 D^{(1)} + (\nu E_4^3+\rho \Delta)\right]$ \\  \bottomrule
\end{tabular}\caption{List of MLDO's of interest}\label{MLDO}
\end{table}

In all the MLDE's the accessory parameter denoted by $\rho$. The $(4,4)$ 
MLDE has two accessory parameters which we denote by $\rho_1$ and $\rho$ 
which we convert into a single parameter by setting $\rho_1=0$ -- this 
was chosen as their solutions are involutive duals to solutions of 
$(4,0)$ MLDEs. All other parameters are determined in terms of the 
exponents at $\tau=i\infty$. Explicit formulae are given in Appendix 
\ref{MLDEw}. The monodromy about the elliptic points are determined by 
the Frobenius exponents at the elliptic points and are given in Table 
\ref{FExp}.

\begin{table}[h]
\centering
\begin{tabular}{c|c|c}
$(n,\ell)$ &  Exponents at $\tau=i$ &  Exponents at $\tau=e^{2\pi i/3}$ \\ \hline
$(4,2)$ &  $[0, 1/2, 1, 3/2]$ &  $[0, 1/3,2/3,4/3]$  \\
$(4,4)$ & $ 0, 1/2, 1, 3/2$ & $0, 1/3,2/3,5/3 $ \\
$(5,2)$ & $[0, 1/2, 1, 3/2,2]$ & $[0, 1/3,2/3,1,5/3]$ \\
$(6,0)$ & $[0, 1/2, 1, 3/2, 2, 5/2]$ & $[0, 1/3,2/3,1,4/3,5/3]$
\end{tabular}
\caption{The Frobenius exponents at the elliptic points for the four 
classes of MLDEs.}\label{FExp}
\end{table}

\subsection{Admissible solutions from the $(3,3)$ MLDE -- the original approach}

The $(3,3)$ MLDE has an accessory parameter and admissible solutions 
using the MMS version of the holomorphic modular bootstrap was studied 
and solved by \cite{Gowdigere:2023xnm}. We summarise their result here.  
The calculation proceeds by inserting the $q$-series expansion of the 
vacuum character into the MLDE. Evaluating this at the leading and 
next-to-leading orders allows all three MLDE parameters (rigid and 
non-rigid) to be expressed entirely in terms of the vacuum index 
$\alpha_0$ (which is proportional to the central charge $c$) and the 
first two Fourier coefficients, $m_1$ and $m_2$.  Evaluating the MLDE at 
the next order in the $q$-expansion yields a polynomial equation linking 
$m_1$, $m_2$, $m_3$, and a scaled central charge variable $N$ (defined 
as $N \equiv -427622160\alpha_0$). Furthermore, substituting the known 
parameters back into the indicial equation produces a cubic equation for 
the characteristic exponents that factorizes into a linear and a 
quadratic piece. For the remaining two conformal weights to be real and 
rational, the discriminant of this quadratic equation must be a perfect 
square, denoted $k^2$ .

Thus a differential equation is transformed into a discrete algebraic 
one. The solution-generating procedure requires simultaneously solving 
these two Diophantine equations for the non-negative integers $(N, m_1, 
m_2, m_3, k)$. For each valid integer set, the corresponding MLDE is 
then solved to higher orders to rigorously verify that all Fourier 
coefficients for the characters remain non-negative integers, thus 
confirming the solution is physically admissible. Fifteen admissible 
solutions were found in this fashion. It is not feasible to use this 
method in our examples and we use the simplification due to KLP 
mentioned in Step 2 of the updated holomorphic modular bootstrap.

\section{Implementing the updated holomorphic bootstrap}

We discuss in detail how the updated holomorphic modular bootstrap works illustrating how 
it works using an example.

\subsection{Determining the allowed exponents modulo one}

Solving for the exponents that lead to admissible solutions of an MLDE 
involves Diophantine equations that are not easy to solve as the number 
of characters increases. However, if we are able to guess the exponents, 
we need to solve only for the accessory parameters, which is a 
relatively easier task. In particular, there is nothing more to do for 
rigid MLDE's. As an intermediate step, using the representation theory 
of $\psl{\BZ_N}$, Kaidi-Lin-Parra-Martinez (KLP) provide a list of 
exponents modulo one for $n\leq 5$\cite{Kaidi:2021ent}. We extend their 
results to the case of $6$ characters.

We briefly discuss the method used by KLP that goes back to 
Eholzer\cite{Eholzer:1994th} (who summarises results in German on the 
representation theory of $\psl{\mathbb{Z}_{p^\lambda}}$ due to Nobs and 
Wolfart\cite{Nobs:1976a,Nobs:1976b}). Fix the number of characters, $n$ 
and then ask for the values $N$ such that $\psl{\mathbb{Z}_N}$ admits a 
$n$-dimensional representation. This is equivalent to asking for 
representations of $\text{SL}(2,\mathbb{Z}_N)$ for which $-I$, that lies 
in the center of $\text{SL}(2,\mathbb{Z}_N)$, acts as the identity. The 
order of the $T$ matrix is then $N$ -- this is the least common multiple 
of the denominators of all exponents.  For $n=6$, we find the possible 
values of $N$ to be:
\[
\begin{split}
(4,6,7,8,9,10,12,14,16,18,20,21,24,28,30,32,36, 40,42,44,48, \qquad \\ 52,56,60,80,
84,96,120,132,140,156,168,240,420)
\end{split}
\]

One can do even better. The allowed exponents modulo one can be 
extracted based on the character table for $\text{PSL}(2, \mathbb{Z}_N)$ 
\footnote{Let $N=\prod_i p_i^{\lambda_i}$ be the prime decomposition of 
$N$. Then, one has $\text{SL}(2,\mathbb{Z}_N)=\otimes_i 
\text{SL}(2,\mathbb{Z}_{p_i^{\lambda_i}})$. This simplifies the problem 
especially when the denominators are large.}. For the allowed 
denominators, the character table is determined and the conjugacy class 
that corresponds to the conjugacy class of the $T$-matrix is chosen 
(there can be multiple possibilities). Then, the conjugacy classes 
corresponding to the powers of modular $T$-matrix, $T^p$ are identified. 
Let $\tilde{\chi}$ denote a character associated with the 
$n$-dimensional representation that is of interest. One performs an 
inverse discrete Fourier transform:
\begin{equation}
    \nu_m = \frac{1}{N} \sum_{p=0}^{N-1} \tilde{\chi}(T^p) e^{-2\pi i m p / N}\ ,
\end{equation}
to obtain $\nu_m$ which is the exact degeneracy of the eigenvalue 
$e^{2\pi i m / N}$ associated with the $n$-dimensional representation. A 
non-zero value for $\nu_m$ implies that the fraction $m/N$ is an allowed 
exponent (modulo 1). The list of allowed denominators and exponents for 
$n\leq 5$ was computed and listed by KLP\cite{Kaidi:2021ent}. The 
character table and the inverse discrete Fourier transform were 
implemented by us in GAP\cite{GAP}. The list of allowed exponents modulo 
one is obtained for $n=6$ and are listed in Table \ref{tab:KLP6}.

\subsection{Admissible solutions for MLDEs with one accessory parameter}

In order to obtain a finite list of allowed exponents after fixing the 
modulo one ambiguity, we focus on obtaining admissible solutions with 
$c_{\text{eff}}\leq 24$. Recall that a priori we do not know if an 
admissible solution is associated with a unitary or non-unitary RCFT. 
One defines
\[
c_{\text{eff}} := (c - h_\text{min})\quad.
\]
Then $-\frac{c_{\text{eff}}}{24}$ is smallest exponent in the list of 
exponents that we consider. Thus, our constraint is equivalent to all 
exponents being $\geq -1$. We will, somewhat loosely, call the character 
associated with the smallest exponent the \textit{identity} character.  
For non-unitary theories, one similarly defines an effective weight 
$h_{\text{eff}}=h-h_\text{min}$. This is called the \textit{unitary 
presentation} of the non-unitary theory\cite{Chandra:2018pjq}.

We focus on the following families of MLDEs: $(4,2)$, $(5,2)$, $(6,0)$. 
We fix one of the accessory parameters in the $(4,4)$ MLDE which then 
behaves like an MLDE with one accessory parameter\footnote{Something 
similar happened for $(3,4)$ MLDEs where the sole accessory parameter 
got fixed to a single value for all admissible 
solutions.\cite{Gowdigere:2023xnm}}. We fix a value of $(n,\ell)$ from 
the list and pick one set of exponents from the finite list. We impose 
the normalization condition $a_{0,0}=1$ -- this is expected to be true 
for non-unitary theories as well\cite{Gannon:2003de}. We then solve for 
the accessory parameter, $\rho$, in terms of the next coefficient, i.e., 
$\mathfrak{m}=a_{0,1}$. This is a linear relationship. Using the MLDE with 
$\rho$ rewritten in terms of $\mathfrak{m}$, we solve for the remaining 
coefficients $a_{0,i}$ for $i\geq2$. We then impose the conditions that 
$\mathfrak{m}$ as well as $a_{0,2}$ and $a_{0,3}$ are non-negative integers. 
Next, we check for additional conditions on $m_1$ that arise from the 
non-negativity of $a_{0,4}$ and $a_{0,5}$. However, we do not go to very 
large values of $\mathfrak{m}$ (we go up to around $10^5-10^6$) as that will require us to 
consider higher terms in the $q$-series to rule them out. In all cases, 
we find a single value of $\mathfrak{m}$ that we suspect might be unique but we 
do not prove this. This in turn fixes the accessory parameter. We then 
check that non-negative integrality holds for the remaining coefficients 
(to order 80, sometimes we go higher) in the identity character. Once 
this holds, we check that the other characters are also admissible to 
the same order. We then deem this solution as \textit{admissible}. We 
illustrate this in detail for one example.

\subsubsection{An ambiguity in the admissible solutions}

Let $\widetilde{\mathbb{X}} = (\tilde\chi_0,\ldots,\tilde\chi_{n-1})^T$ 
be a of rank $n$ solution. While solving the MLDE, it is conventional to 
choose the normalizations of the $\widetilde{\chi}_i$ for $i\neq0$ to 
take its minimal possible value. However, this minimal value might not 
be the correct one. It could be that the RCFT is associated with
\[
\mathbb{X} = \begin{pmatrix} \tilde\chi_0 \\ \mathbf{y_1}\  \tilde{\chi}_1 \\ \vdots \\ \mathbf{y_{n-1}}\ 
\tilde\chi_{n-1}\end{pmatrix}\ ,
\]
where the $(\mathbf{y_1},\ldots,\mathbf{y_{n-1}})\in 
\mathbb{Z}_{>0}^{n-1}$ to account for the ambiguity. Suppose the 
diagonal modular invariant toroidal partition function is given by
$$
Z= |\chi_0|^2 + \sum_{i=1}^{n-1} m_i\ |\chi_i|^2\ =: \bar{\mathbb{X}}^T 
\cdot M \cdot \mathbb{X}.
$$
It is easy to see that the multiplicities $\tilde{m}_i$ associated with 
$\widetilde{\mathbb{X}}$ are then given by
\[
\tilde{m}_i = \mathbf{y_i}^2 \  m_i\ .
\]
This is an ambiguity that needs to be fixed. One solution, that we 
typically use, is to require the $m_i$ to be square-free. Any square 
that is a factor of $m_i$ can be absorbed in the $\mathbf{y_i}$. The 
$S$-matrix also changes under the rescaling. One has
\[
\tilde{S}_{jk} = 
\left(\prod_{i=1}^{n-1}(\mathbf{y_i})^{\delta_{ik}-\delta_{ji}}\right) \ 
S_{jk} \ .
\]

\subsubsection{Admissible solutions for \textbf{(4,2)} MLDE}
The (4,2) MLDE is given by
\begin{equation}\label{eq:42MDE}
\left[ D^{(4)}+ \frac13 \frac{E_6}{E_4}\, D^{(3)} + \kappa E_4\, D^{(2)}  + \mu E_6\, D +(\nu E_4^2+\rho \frac{\Delta}{E_4})\right]  f(\tau)=\ 0~,
\end{equation}
where $\rho$ is the accessory parameter. As a working example, consider 
one set of exponents modulo 1 from the KLP list i.e., 
($\frac{7}{9},\frac{1}{9},\frac{1}{3},\frac{4}{9}$). One choice that 
fixes that mod 1 ambiguity is $\alpha_i:\,( 
-\frac{2}{9},\frac{1}{3},\frac{1}{9},\frac{4}{9})$. This corresponds to 
central charge $c=\frac{16}{3}$ and weights 
$H=(0,\frac{5}{9},\frac{1}{3},\frac{2}{3})$. For the given exponents, 
the parameters of the MLDE that are fixed by the exponents are 
\(\kappa=-\frac{11}{108},\, 
\mu=\frac{163}{5832},\,\nu=-\frac{8}{2187}\). Using the MLDE, we express 
the first few terms of the identity character in terms of $m_1$. We find

\begin{align*}\rho &= -\frac{8}{81} \left(\mathfrak{m}+110\right)\ ,\\
a_{0,2}&=\frac1{65}\left(\mathfrak{m}^2+465 \mathfrak{m}-1064\right)\ ,\\
a_{0,3}&= \frac{1}{30030}\left(\mathfrak{m}^3-6777 \mathfrak{m}^2+1002006 \mathfrak{m}+30736288\right) \ .
\end{align*}
We obtain an admissible solution when $\mathfrak{m}=53$. We also found 
no admissible solutions $\mathfrak{m}\leq 356153$ from studying the 
first five terms. The value of $\rho$ is $-\frac{1304}{81}$ for 
$\mathfrak{m}=53$.
\begin{equation}\label{42soln}
\widetilde{\mathbb{X}}=\left(\begin{smallmatrix}
q^{-2/9}(1+53 q+ 406 q^2+2163q^3+8824q^4+\cdots)\\
y_1\ q^{1/3} (1+13 q+ 77q^2+351q^3+1313q^4+\cdots) \\
y_2\ q^{1/9}\ (1+28 q+ 159 q^2+813q^3+3079q^4+\cdots) \\
 y_3\ q^{4/9}  (25+271 q+ 1568 q^2+6908q^3+25544q^4+\cdots) 
\end{smallmatrix}\right)
\end{equation}

The minimal normalized solution given above as obtained from MLDE has an 
ambiguity given by the three positive non-zero integers $(y_1,y_2,y_3)$ 
as shown above. We will fix it using the $S$-matrix. We then obtain a 
solution with the correct
\[
\mathbb{X}=\left(\begin{smallmatrix}
q^{-2/9}(1+53 q+ 406 q^2+2163q^3+8824q^4+\cdots)\\
\mathbf{9}\ q^{1/3}(1+13 q+ 77q^2+351q^3+1313q^4+\cdots) \\
q^{1/9} (1+28 q+ 159 q^2+813q^3+3079q^4+\cdots) \\
 q^{4/9} (25+271 q+ 1568 q^2+6908q^3+25544q^4+\cdots) 
\end{smallmatrix}\right)
\]

\subsection{From an admissible solution to a tenable solution}

We then compute the numerical $S$-matrix using the method discussed in 
ref. \cite{Govindarajan:2026frs}. We rewrite the $(4,2)$ MLDE in the 
coordinate $w=\frac{1728}{J(\tau)}$. In this coordinate, $\tau=i\infty$ 
gets mapped to $w=0$ and $\tau=i$ gets mapped to $w=1$. The monodromy 
about $w=0$ is the $T$ matrix and the monodromy about $w=1$ is the 
$S$-matrix.
\begin{equation*}
\begin{split}
\left[\partial_w^4+\left(\tfrac{16/3}{w}+\tfrac{3}{w-1}\right)\partial^3_w+\left(\tfrac{185/36+\kappa}{w^2}+\tfrac{29/3-\kappa}{w(w-1)}+\tfrac{3/4}{(w-1)^2}\right)\partial^2_w+\right.  
\hspace{1.7in} &\\ \left. 
\left(\tfrac{25/54+5\kappa/6+\mu}{w^3}+\tfrac{35/9-\kappa/3-\mu}{w^2(w-1)}+\tfrac{5/6+\kappa/2}{w(w-1)^2}\right)\partial_w+\tfrac{\nu}{w^4(w-1)^4} 
+\tfrac{\rho}{1728w^3(w-1)^2}\right]f(\tau)\, &=0~.
\end{split}
\end{equation*}
The MLDEs of interest in this coordinate are given in Appendix 
\ref{MLDEw} where one we give explicit expressions for the non-accessory 
parameters in terms of the exponents about $w=0$.

Let $\mathcal{B}_0(w)$ denote the basis given by the Frobenius power 
series solution and the local monodromy $M_0$ is the $T$ matrix. 
Similarly, let $\mathcal{B}_1(u)$ denote the basis given by the 
Frobenius power series about $u:=(1-w)=0$ with monodromy $M_1$.  Let 
$C$, as defined below, denote the connection matrix that connects the 
two bases.
\[
B_1(1-w_0) = C \cdot B_0(w_0)\ ,
\]
where $w_0$ is a regular point that we choose to be $\frac12+\epsilon$, 
with $\epsilon$ assumed to be small with magnitude less than half. We then 
determine $C$ numerically by matching the solutions to order 
$\epsilon^{(n-1)}$. It is important that we take sufficiently large 
number of terms in the Frobenius power series to improve the numerical 
accuracy of the connection matrix. The numerical estimate for the 
$S$-matrix is given by
\[
S = C^{-1}\cdot M_1 \cdot C \ .
\]
\subsubsection*{Illustrating using our working example}
Solving the MLDE in $w$ and $u=(1-w)$ coordinates, we look for and 
obtain solutions of the form: ($\lim_{w\rightarrow0} \frac{w}{1728}=q$)
\begin{align*}
\mathbb{B}_0(w)&=\begin{pmatrix}
(\frac{w}{1728})^{-2/9}\Big[1+ O(w)\Big] \\
(\frac{w}{1728})^{1/3}\Big[9+ O(w)\Big] \\
(\frac{w}{1728})^{1/9}\Big[1+ O(w)\Big] \\
(\frac{w}{1728})^{4/9}\Big[25+ O(w)\Big] 
\end{pmatrix}\ , \ 
\mathbb{B}_1(u)=\begin{pmatrix}
(1+ O(u) )\\
u^{1/2} (1+ O(u)) \\
u (1+ O(u) )\\
u^{3/2}(1+ O(u) )
\end{pmatrix} \ .
\end{align*}
Note that we have chosen $y_1=9$ and $y_2=y_3=1$ in the basis 
$\mathcal{B}_0$. The normalizations for the basis $B_{1}$ are not so 
crucial as they do not change the $S$-matrix. Also, the exponents at 
$u=0$ are as given in Table \ref{FExp}.

The local monodromy matrices are easy to obtain. We see that
\[
T=M_0 = \text{Diag}\left(e^{-4\pi i/9},e^{2\pi i/3},e^{2\pi i/9}, e^{8\pi i/9}\right)\quad,\quad
M_1 = \text{Diag}\left(1,-1,1,-1\right)\ .
\]
The connection matrix is given by
\[
C=\begin{pmatrix}
0.02484 & 2.169 & 0.8520 & -1.292 \\
 0.001747 & -7.181 & 9.281 & 2.102 \\
 0.5810 & -8.237 & -3.360 & 5.458 \\
 2.070 & 54.20 & -76.05 & -19.78
\end{pmatrix}
\]

The $S$-matrix for our working example with the normalizations $y_1=9$ 
and $y_2=y_3=1$ in Eq. \eqref{42soln} is found to be
\[
S_\text{num}=
\begin{pmatrix}
 0.5107 & 1.000 & 0.1158 & 0.6265 \\
 0.3333 & -2.98\times10^{-10} & 0.3333 & -0.3333 \\
 0.1158 & 1.00 & -0.6265 & -0.5107 \\
 0.6265 & -1.000 & -0.5107 & 0.1158
\end{pmatrix}
\]
One of the results of \cite{Govindarajan:2026frs} is that one can 
convert the numerical $S$-matrix to an exact one. Let $\bar{S}$ denote 
the normalized $S$-matrix defined by $S/S_{00}$. We obtain
\begin{equation}\label{42Smatrix}
\bar{S} = 
\begin{pmatrix}
1.000 & 1.958 & 0.2267 & 1.227 \\
 0.6527 & -5.837\times10^{-10} & 0.6527 & -0.6527 \\
 0.2267 & 1.958 & -1.227 & -1.000 \\
 1.227 & -1.958 & -1.000 & 0.2267
\end{pmatrix}
\end{equation}
Then, the entries of the normalized $S$ matrix lie in the integer ring, $\mathbb{Z}(\zeta_{\bar{N}})$,
of the field $\mathbb{Q}(\zeta_{\bar{N}})$, where $\bar{N}$ is the order 
of the normalized $T$ matrix, i.e., 
$T/T_{00}$\cite{DeBoer:1990em,Ng:2007}. In our working example, 
$\bar{N}=9$. The exact normalized $S$-matrix is given by
\begin{equation}
\bar{S}= \left(\begin{smallmatrix}
    1 &  3 \gamma_1 & \gamma_2 & 1+ \gamma_2 \\
    \gamma_1 & 0 & \gamma_1 & -\gamma_1 \\
    \gamma_2 &  3 \gamma_1 & -1-\gamma_2 & -1 \\
       1+\gamma_2 & - 3 \gamma_1 & -1 & \gamma_2
\end{smallmatrix}\right)
\end{equation}
where $\gamma_1 = (1- c_9^2)$ and $\gamma_2=(-2 \gamma_1 + c_9^1)$. One 
can also show that $S_{00}=\frac13 c_9^1$. The modular invariant 
partition function is then
\[
Z = |\chi_0|^2 + 3 |\chi_1|^2 + |\chi_2|^2 +|\chi_3|^2 \ .
\]

\noindent \textbf{Remark:} We found that for large values of $\bar{N}$, 
the simple method used in \cite{Govindarajan:2026frs} finds local 
minima. However, there is a simple implementation of the 
Lenstra–Lenstra–Lovász lattice basis reduction algorithm in PARI/GP that 
works better\cite{PARI2}. Here is an illustration of the how we used it.
\begin{verbatim}
? lindep([1,2*cos(2*Pi/9),2*cos(4*Pi/9),0.6527],3)
%1 = [-1, 0, 1, 1]~
\end{verbatim}
The input to the command `lindep' is a basis $[1,c_9^1,c_9^2,0,6527]$ 
followed by the accuracy which we chose to be 3. The output shows that 
$-1+ c_9^2+0.6527\sim 0$. Needless, to say we go to much higher accuracy 
in our computations. The bases tend to be larger as $\bar{N}$ increases. 
In our examples, the ones with $\bar{N}=35,80$ forced us to use 
PARI/GP\footnote{We thank Boris Pioline who emphasized the need for 
better algorithms.}.

In order to compute the fusion matrices we need to figure out the label 
of the identity operator. The first guess is that the zeroth label 
corresponds to the identity -- that would be true in a unitary RCFT. 
This leads to fusion coefficients with integral coefficients, but some 
are negative. The second possibility is the operator labeled 2 as it 
also has degeneracy one. This leads to fusion rules that have positive 
integral coefficients. Thus, we have a non-unitary RCFT (if it is one).  
Recall that operator labeled 1 has multiplicity 3 and we need to resolve 
the $S$-matrix into a $6\times 6$ matrix\footnote{We do not provide the 
resolution of this $S$-matrix as that is not the intent of this paper. 
However, in all instances where we either identify it with a known MTC 
(only for unitary cases) or a known RCFT, the resolved S-matrix is 
known.} and then require fusion coefficients to be either $0$ or $1$. So 
we have
\[
c=-\frac{8}{3}\text{ and }H=\left(-\frac{1}{3},\left(\frac29\right)^3,0,\frac13\right)\ ,
\]
where the superscript indicates the multiplicity.  We label this 
solution as a non-unitary tenable solution.

\section{Other approaches to fixing the normalization of admissible solutions}

The ambiguity in determining $y_i$ can be fixed using other methods. We 
discuss three of them: (i) Generalized GHM duality, (ii) Using the Hecke 
operator method, and (iii) Involutive Duality.  All three methods map 
known solutions to new ones. If we can map our admissible solutions to 
known ones using any one of these methods, a natural value is obtained 
for $y_i$, the multiplicities $m_i$ and the $S$-matrix. The method is 
somewhat ad hoc and may not always work. However, we have checked our 
results using these methods in a fraction of the admissible solutions we 
found. It appears to be consistent with the method discussed in the 
previous section for these solutions.

\subsection{Method 1: Finding a GHM dual}

Let $\mathcal{G}$ be a meromorphic CFT with central charge $8k$ and 
character $\chi_0(\tau)$. Let $\mathbb{X}$ be the VVMF's of an RCFT 
$\mathcal{H}$ with $n$ characters, Wronskian index $\ell$ and 
multiplicities $m_i$ which we identify as the entries in the diagonal 
matrix $M$. The generalized Gaberdiel-Hampapura-Mukhi (GHM) coset 
dual\cite{Gaberdiel:2016zke}, $\mathcal{C}$, whose characters form a 
VVMF, $\tilde{\mathbb{X}}$ which satisfies the following condition:
\[
\chi_0=\tilde{\mathbb{X}}^T \cdot M \cdot \mathbb{X}  \ .
\]
One has $\chi_0(\tau)=J(\tau)^{1/3}$ for $k=1$; 
$\chi_0(\tau)=J(\tau)^{2/3}$ for $k=2$ and 
$\chi_0(\tau)=J(\tau)-744+\mathcal{N}$ for $k=3$($\mathcal{N}\geq0$).

The resulting coset $\mathcal{C}$ contains a Virasoro algebra with a 
central charge given by the difference $c_{\mathcal{C}} = 
c_{\mathcal{G}} - c_{\mathcal{H}}=8k-c_{\mathcal{H}}$, and it leads to a 
well-defined CFT. The number of characters $n$ of the denominator theory 
$\mathcal{H}$ and the coset $\mathcal{C}$ must be equal. Furthermore, if 
$h_i$ and $\tilde{h}_i$ represent the conformal dimensions of the 
non-vacuum primaries in $\mathcal{H}$ and $\mathcal{C}$ respectively, 
they are required to pair up to integers: $h_i + \tilde{h}_i = 
(h_{sum})_i \in \mathbb{N}$. The Wronskian index $\ell\ (\equiv 
\ell_\mathcal{H})$ of the denominator theory and $\tilde{\ell}(\equiv 
\ell_\mathcal{C})$ of the coset theory are algebraically constrained by 
the formula ;
\begin{equation}
    \ell + \tilde{\ell} = n^2 + (2k - 1)n - 6 \sum_{i=1}^{n-1} (h_{sum})_i.
\end{equation}
The relevant pairings in the one accessory parameter examples can be 
summarized as:
\begin{table}[h]
    \centering
    \begin{tabular}{|ccc|ccc|ccc|ccc|}
    \toprule
  \multicolumn{3}{|c|}{$(4,2)$ }& \multicolumn{3}{c|}{$(4,4)$ } & \multicolumn{3}{c|}{$(5,2)$}& \multicolumn{3}{c|}{$(6,0)$}\\ \midrule
     $k$ & $\tilde{\ell}$ & $\sum h_{sum}$ & $k$ & $\tilde{\ell}$ &$\sum h_{sum}$  & $k$ & $\tilde{\ell}$ & $\sum h_{sum}$ & $k$ & $\tilde{\ell}$ &$\sum h_{sum}$ \\
     \midrule
     1 & 0 & 3 & 2 & 0 &4  & 1& 4& 4& 1& 0&7\\ \midrule
    2 & 2 & 4 & 3 & 2 &5  & 2& 2& 6& 2& 0&9\\ \midrule
    3 & 4 & 5 & 4 & 4 &6  & 3& 0& 8& 3& 0&11\\
     \bottomrule
    \end{tabular}
    \caption{Possible coset pairs for theories with one accessory parameter}
    \label{tab:GHM}
\end{table}
\subsubsection*{Applying this to the working example}

The $(4,2)$ tenable solution is the GHM dual of $(4,0)$ solution with 
$\ceff=\frac{8}{3}$, $\heff=(0,\frac{4}{9},\frac{2}{3},\frac{1}{3})$. 
The $(4,0)$ theory has been found by KLP\cite[see Table 
5]{Kaidi:2021ent} and has been discussed by \cite{Duan:2022ltz}. One has 
the bilinear relation
 \begin{equation*}
    J(\tau) ^{1/3}= \tilde{\chi_0}(\tau) \chi_0(\tau)  +3\tilde{\chi}_{\frac{5}{9}}(\tau) \chi_{\frac{4}{9}}(\tau) +\tilde{\chi}_{\frac{1}{3}}(\tau) \chi_{\frac{2}{3}}(\tau) + \tilde{\chi}_{\frac{2}{3}}(\tau) \chi_{\frac{1}{3}}(\tau) .
\end{equation*}
The multiplicity $m_1=3$ agrees with our earlier conclusion. 
Furthermore, the $(4,0)$ theory has the same $S$-matrix that we 
determined earlier.

We also find a $k=2$ GHM dual that is also another $(4,2)$ tenable 
solution that we obtained (see Table \ref{tab:42}). It has 
$\ceff=\frac{32}{3}$, $\heff=(0,\frac{4}{9},\frac{2}{3},\frac{4}{3})$. 
The bilinear relation is
 \[
  J(\tau) ^{2/3}= \tilde{\chi}_0(\tau) \chi_0(\tau) + +3\tilde{\chi}_{\frac{5}{9}}(\tau) \chi_{\frac{4}{9}}(\tau) +\tilde{\chi}_{\frac{1}{3}}(\tau) \chi_{\frac{2}{3}}(\tau) + \tilde{\chi}_{\frac{2}{3}}(\tau) \chi_{\frac{4}{3}}(\tau) 
 \]
There is a $(4,4)$ theory that has $\ceff=\frac{56}{3}$, $\heff=(0, 
\frac{13}{9}, \frac{2}{3}, \frac{4}{3})$ that is a GHM dual with $k=3$. 
Again, this appears as a tenable solution in Table \ref{tab:44}. The 
bilinear relations is
\begin{align*}
   J(\tau)-508&= \tilde{\chi}_0(\tau) \chi_0(\tau) + +3\tilde{\chi}_{\frac{5}{9}}(\tau) \chi_{\frac{13}{9}}(\tau) +\tilde{\chi}_{\frac{1}{3}}(\tau) \chi_{\frac{2}{3}}(\tau) + \tilde{\chi}_{\frac{2}{3}}(\tau) \chi_{\frac{1}{3}}(\tau)\ .
\end{align*}

\subsection{Obtaining tenable $(5,0)$ solutions as GHM duals of tenable $(5,2)$ solutions}

Table \ref{tab:GHM} shows that $(5,2)$ solutions are dual to other $(5,2)$ solutions.  We can indeed  verify that the tenable $(5,2)$ examples given in \ref{tab:52} are GHM duals to each other with $k=2$. Similarly, tenable $(5,0)$ solutions  that are $k=3$ GHM duals of $(5,2)$ tenable solutions are given in Table \ref{tab:50tenable}. This is the complete list of tenable $(5,0)$ solutions and is consistent with the results given in \cite{Kaidi:2021ent} who did not obtain the $S$-matrices. Further, the $S$-matrices and degeneracies are consistent with those given in \cite{Duan:2022ltz}. Needless to say, we also derived the same results using the updated bootstrap method introduced in this paper.
\begin{table}[h]
\begin{tabular}{l||l|l}\toprule
$(5,2)$  $\ceff,\heff$ & $(5,0)$ $\ceff,\heff$ & $(\mathfrak{m},a_{i,0})$ \\ \midrule
     & $\frac{104}{5}$, $(0,\frac{9}{5}, \frac{8}{5}, \frac{7}{5}, \frac{6}{5})$ & $(0,\mathbf{25}\cdot728,\mathbf{25}\cdot91,2808,\mathbf{5}\cdot52$)\\
  $\frac{16}{5}$, $(0,\frac{1}{5}, \frac{2}{5}, \frac{3}{5}, \frac{4}{5})$   &$\frac{104}{5}$, $(0,\frac{9}{5}, \frac{3}{5}, \frac{12}{5}, \frac{6}{5})$ & $(208,17576,26,456976,676)$\\
       &$\frac{104}{5}$, $(0,\frac{4}{5}, \frac{8}{5}, \frac{7}{5}, \frac{11}{5})$& $(520,\mathbf{5}\cdot26,\mathbf{5}\cdot429,2288,\mathbf{125}\cdot624)$ \\ \midrule
   $\frac{64}{5}$, $(0,\frac{4}{5}, \frac{8}{5}, \frac{2}{5}, \frac{6}{5})$ &  $\frac{56}{5}$, $(0,\frac{6}{5}, \frac{2}{5}, \frac{8}{5}, \frac{4}{5})$ & $(56,343,7,2401,49)$ \\
$\frac{64}{5}$,$(0,\frac{4}{5}, \frac{3}{5}, \frac{7}{5}, \frac{6}{5} )$      & $\frac{56}{5} ,(0,\frac{6}{5}, \frac{7}{5}, \frac{3}{5}, \frac{4}{5})$ &$(28,\mathbf{25}\cdot14,\mathbf{25}\cdot12,28,\mathbf{5}\cdot7)$ \\ \midrule
 $\frac{16}{11}$, $(0,\frac{1}{11}, \frac{2}{11}, \frac{6}{11}, \frac{9}{11})$ & $\frac{248}{11}$, $(0,\frac{10}{11},\frac{20}{11},\frac{16}{11},\frac{24}{11})$ & $(248,248,30380,3875,147250)$\\ 
     & $\frac{248}{11}$, $(0,\frac{21}{11},\frac{20}{11},\frac{16}{11},\frac{13}{11})$ & $(0,61256,30628,4123,248)$\\ \midrule
 $\frac{160}{11}$,  $(0,\frac{9}{11}, \frac{10}{11}, \frac{13}{11}, \frac{16}{11})$ & $\frac{104}{11}$,$(0,\frac{13}{11}, \frac{12}{11}, \frac{9}{11}, \frac{6}{11})$ & $(52,324,273,52,26)$ \\ \bottomrule
\end{tabular}
\caption{We list all tenable $(5,0)$ solutions with $\ceff\leq 24$ as GHM duals of $(5,2)$ tenable solutions. The $S$-matrices and degeneracies are identical to the $(5,2)$ duals.}
\label{tab:50tenable}
\end{table}
A few more of our $(5,2)$ solutions appear as $k=3$ GHM duals of $(5,0)$ solutions obtained using the Hecke operator method (that we discuss next)\cite{Duan:2022ltz}. We can also see that non-tenable examples 5.14 and 11.2 (in Table \ref{tab:52}) are $k=3$ GHM duals to $\mathcal{M}(5,2)^{\otimes4}$  and $\mathcal{M}(11,2)$ respectively.

\subsection{Method 2: The Hecke operator method}

Harvey and Wu\cite{Harvey:2018rdc} generalized Hecke operators to VVMFs. 
Consider a VVMF for which the kernel of the representation is a level 
$N$ (the conductor) subgroup. The Hecke operator (for some prime $p\nmid 
N$) is defined by
\begin{equation*}
   ( T_p\chi)_{a}(\tau)=\sum_{b}v_{ab}(\sigma_p)\chi_b(p\tau)+\sum_{k=0}^{p-1}\chi_a(\frac{\tau+kN}{p})
\end{equation*}
Here $\sigma_p$ can be chosen as 
$T^{\bar{p}}S^{-1}T^pST^pST^{\bar{p}}S$, where $p\bar{p}=1\mod N$. For 
VVMFs arising from RCFTs, Bantay has shown that $N$ is the order of the 
$T$-matrix\cite{Bantay:2001ni}. Acting on a VVMF associated with an RCFT 
by this Hecke operator can lead to new admissible solutions.

Duan et al. used the Hecke operator method to systematically construct 
several examples of theories with up to seven 
characters\cite{Duan:2022ltz}. In the process, they provide several 
examples that realize the exponents (modulo one) of 
KLP\cite{Kaidi:2021ent}. They also provide the $S$-matrices and 
degeneracies for their examples. All their admissible $(5,2)$ and 
$(6,0)$ examples appear in our Tables. More generally, we find Hecke 
operators mapping $(4,2)$ to $(4,4)$, $(5,0)$ to $(5,2)$ and $(6,0)$ to 
$(6,0)$.

For our tenable $(4,2)$ example, the conductor is $N=9$. For $p=2$, the 
multiplier associated with $\sigma_p$ is
\[
v(\sigma_2)=\begin{pmatrix}
     0 & 0 & 0 & -1 \\
 0 & 1 & 0 & 0 \\
 1 & 0 & 0 & 0 \\
 0 & 0 & -1  & 0 \\
\end{pmatrix}\ ,
\]
which is a signed permutation matrix.
The $T_2$ image can be determined to be 
\[
T_2\chi=\begin{pmatrix}
 q^{-4/9}(1+ 56q+1679 q^2+ 21100q^3+\cdots)\\
    243q^{2/3}(1+ 26q+ 323q^2+ 2704q^3+\cdots)\\
    q^{2/9}(49+ 3136q+ 51060q^2+497696 q^3+\cdots)\\
    q^{-1/9}(2+ 787q+17648 q^2+ 197383q^3+\cdots)
\end{pmatrix}
\]
This corresponds to the $(4,4)$ solution with $\ceff=\frac{32}{3}$ , 
$\heff=(0,\frac{10}{9}, \frac{1}{3}, \frac{2}{3})$ which is a tenable 
solution in Table \ref{tab:44}. Similarly one can find such relations in 
$(6,0)$ solutions. Consider $\ceff=\frac{5}{7}$ , 
$\heff=(0,\frac{1}{28},\frac{5}{28},\frac{2}{7},\frac{10}{7},\frac{3}{4})$ 
tenable solution from \ref{tab:60} with conductor $N=168$.  For $p=11$, 
the multiplier associated with $\sigma_p$ is
\[
v(\sigma_{11})=\begin{pmatrix}
 0 & 0 & 0 & 0 & 1 & 0 \\
 0 & 0 & 0 & 0 & 0 & -1 \\
 0 & 1 & 0 & 0 & 0 & 0 \\
 -1 & 0 & 0 & 0 & 0 & 0 \\
 0 & 0 & 0 & 1 & 0 & 0 \\
 0 & 0 & 1 & 0 & 0 & 0 \\
\end{pmatrix}\ ,
\]
from which the  $T_{11}$ image is determined to be 
\[T_{11}\chi=\begin{pmatrix}
 q^{-55/168}(1+ 55q+ 880q^2+ 7315q^3+\cdots)\\
 2q^{11/168}(5+ 16q+ 1925q^2+ 13706q^3+\cdots)\\
  2q^{155/168}(66+ 770q+ 5885q^2+ 33055q^3+\cdots)\\
   11q^{65/168}(4+ 90q+ 850q^2+ 5500q^3+\cdots)\\
  2q^{107/168}(55+ 919q+ 7645q^2+ 46420q^3+\cdots)\\
   q^{137/168}(165+ 2211q+ 17280q^2+ 100155q^3+\cdots)\\
\end{pmatrix}\]
corresponding to $\ceff=\frac{55}{7}$ , 
$\heff=(0,\frac{11}{28},\frac{5}{4},\frac{5}{7},\frac{27}{7},\frac{8}{7})$ 
tenable solution from Table \ref{tab:60}.

\subsection{Method 3: Involutive dual}

Given a weakly holomorphic VVMF, $\mathbb{X}$, of rank $n$, the theory 
of VVMFs due to Bantay and Gannon\cite{Bantay:2007zz,Gannon:2013jua} 
lets one construct a basis, 
$\Xi=(\mathbb{X},\mathbb{Y}_1,\ldots,\mathbb{Y}_{n-1})$, for VVMFs that 
share the same multiplier. This is applicable to the case when 
$\mathbb{X}$ arises from the characters of an RCFT. It has been shown 
that the new VVMFs i.e., $\mathbb{Y}_i$ ($i=1,\ldots,(n-1)$) are 
generically quasi-characters, though sometimes, they correspond to 
another RCFT\cite{Govindarajan:2025rgh,Govindarajan:2025jlq}. Another 
interesting construction from their work is the duality that relates one 
basis $\Xi$ to another $\Xi^\vee$. This is defined as follows:
\begin{equation}
\Xi^\vee(\tau) := \frac{E_4(\tau)^2 E_6(\tau)}{\Delta(\tau)^{7/6}} \left(\Xi(\tau)^T\right)^{-1}\ .
\label{BGdual}
\end{equation}
When $\Xi$ arises from a solution of type $(n,\ell)$, the dual solution 
is of type $(n,\ell^\vee)$, where
\[
\ell^\vee = (n-5)(n-1)+7 -\ell\ .
\]
This involutive duality relates the $(3,0)$ solutions to $(3,3)$ 
solutions and was used to determine the, then unknown, $S$-matrices for 
the 15 admissible solutions found in \cite{Gowdigere:2023xnm}. We find 
the following involutive pairs:
\[
(4,0) \leftrightarrow (4,4) \quad,\quad (4,2) \leftrightarrow (4,2)\quad,\quad (5,2) \leftrightarrow (5,5)\quad,\quad (6,0) \leftrightarrow (6,12)\ .
\]
As mentioned earlier, the $(4,4)$ admissible solutions that are obtained as involutive duals of $(4,0)$ admissible solutions come from $(4,4)$ MLDE with one of the accessory parameters being zero. Consider the example where $(4,4)$ tenable solution with $\ceff=\frac{113}{5}$, $\heff=(0,\frac{4}{5},\frac{7}{4},\frac{31}{20})$. It appears as the involutive dual to the $(4,0)$ RCFT with $\ceff=\frac{3}{5}$, $\heff=(0,\frac{4}{5},\frac{1}{4},\frac{1}{20})$ -- it is the non-unitary minimal model $\mathcal{M}(5,3)$.  It is also the GHM coset with $c_{\mathcal{G}}=24$ with $(4,2)$ theory $\ceff=\frac{7}{5}$, $\heff=(0,\frac{1}{5},\frac{1}{4},\frac{9}{20})$ ($A_{1,1}\otimes \mathcal{M}(5,2)$). The degeneracy and $S$-matrix obtained in these multiple ways indeed agree with the direct calculation.

The involutive dual of our working example gives two different tenable 
$(4,2)$ solutions: (i) $\ceff=\frac{40}{3}$, 
$\heff=(0,\frac{2}{3},\frac{8}{9},\frac{4}{3} )$ and (ii) 
$\ceff=\frac{20}{3}$, $\heff=(0,\frac{1}{3},\frac{2}{3},\frac{7}{9})$. 
It is important to note that in the presence of multiplicities, 
determining the degeneracy should be done carefully.

\section{Discussion and Concluding Remarks}

In this paper, we have provided an important update to the holomorphic 
modular bootstrap, that has enabled us to directly go to what we call 
tenable solutions. For such solutions, we are able to provide the 
$S$-matrix, fusion rules and multiplicities. As an illustration, we have 
considered MLDEs with one accessory parameter and thereby classifying 
all such theories up to six characters with $\ceff\leq 24$.

We have considered $(4,4)$ theories where one of the accessory 
parameters vanishes. Is it possible for solutions to exist when the 
accessory parameter, call it $\rho_1$, is non-vanishing. This accessory 
parameter changes the exponents at the elliptic point $\tau=2\pi/3$. 
Recall that the exponents when $\rho_1=0$ were 
$(0,\frac13,\frac23,\frac53)$ consistent with the fact that the 
monodromy about the elliptic point, $U$, has eigenvalues given by cube 
roots of unity. The indicial equation with $\rho_1\neq0$ takes the form
\[
\lambda (\lambda-\tfrac{1}{3})(9\lambda^2-21 \lambda + (10-9\rho_1))\ . 
\]
Two of the exponents remain unchanged. However, for arbitrary values of 
$\rho_1$, the other two roots will not equal $\frac23\mod 1$. Setting 
$\rho_1=k(k+1)$ with $k\in \mathbb{Z}_{\geq0}$, the other two exponents 
will become
\[
\frac23-k \quad, \quad \frac53+k\ .
\]
Thus, the non-negative integer $k$ parametrizes the allowed values of 
$\rho_1$. It is of interest to see if there are any $(4,4)$ solutions 
for $k>0$.

Entry $13.3$ in our list of $(6,0)$ admissible solutions is of IVOA 
type\cite{Chandra:2018pjq} appears in list of six character theories 
obtained by Duan et al.\cite[see sec. 7.1]{Duan:2022ltz}. They call this 
theory as $E_{7\frac12,2}$ as it satisfies several conditions expected 
of such a theory. In particular, the central charge at level $k$ was 
guessed to be $c_k=190k/(k+24)$ and the degeneracies were noted to equal 
the dimensions of `representations' of $E_{7\frac12}$\cite{Lee:2023owa}. 
Several, not all, of the IVOA's in our $(4,2)$ list arise from tensor 
product of $E_{7\frac12,1}$ with other two-character theories.

In our list of exponents for the six character solutions, we only found 
examples with Wronskian index $\ell=0,3$. We did not find any with 
Wronskian index $\ell=2$. This is a consequence of the irreducibility 
constraint while computing the exponents. It appears that there are none 
with $\ceff\leq 24$. However, $(6,2)$ theories that violate this 
irreducibility exist. For instance, the RCFT associated with the affine 
algebra $A_{2,3}$ has $c=4$ has six characters and ten primaries. The 
central charge is $c=4$ and the conformal weights are $H=(0, 
2/9,1/2,5/9,8/9,1)$. This theory has Wronskian index $\ell=2$. We can 
see that two of the exponents differ by an integer. Further, the 
characters $(\chi_0+2\chi_1,\chi_{1/2})$ are the two characters of 
$D_{4,1}$ RCFT and thus satisfy a $(2,0)$ MLDE. Here we are labelling 
the characters of $A_{2,1}$ by their conformal weight. The characters 
satisfy the $(6,2)$ MLDE
\begin{multline}
\label{eq:62MDE}
\left[ D^{(6)} + \tfrac13 \tfrac{E_6}{E_4} + \mu_2 E_4  D^{(4)}+ \mu_3 E_6 D^{(3)} \right.\\ + (\mu_4 E_4^2 +\rho_1 \tfrac{\Delta}{E_4}) D^{(2)} 
\left. + (\mu_5 E_4E_6 )D^{(1)} + \nu E_4^3\right]  f(\tau)\,= \, 0~.
\end{multline}
with $\mu_i$ and $\nu$ determined by the exponents as given in Eq. 
\eqref{n6pars}. The lone accessory parameter $\rho_1=-(557/17496)$. The 
other potential accessory parameter which appears in the $(6,0)$ MLDE is 
vanishing here. Two more $(4,2)$ theories, obtained as a tensor product 
of two two-character theories, $E_{7,1}\otimes A_{1,1}$ (with $\ceff=8$ 
and $\heff=(0,1/4,3/4,1)$) and $E_{7\frac12,1}\otimes \mathcal{M}(5,2)$ 
(with $\ceff=8$ and $\heff=(0,1/5,4/5,1)$) do not appear in our list of 
$(4,2)$ admissible solutions for the same reason. Similarly, the $(6,0)$ 
theory corresponding to $G_{2,3}$ with central charge $c=6$ and weights 
identical to those for $A_{2,3}$ given earlier doesn't appear in our 
list of $(6,0)$ theories (see Table \ref{tab:60}).

In the context of the theory of VVMFs developed largely by Bantay and 
Gannon, the matrix MLDE that appeared in \cite{Gannon:2013jua}, requires 
one to fix exponents $\Lambda=(\lambda_0,\ldots,\lambda_{n-1})$ exactly 
without any modulo one ambiguity. In the context of VVMFs that arise 
from RCFTs, one has $\lambda_i=\alpha_i\mod 1$. In earlier 
work\cite{Govindarajan:2025rgh,Govindarajan:2025jlq}, the following 
choice worked:
\begin{equation}\label{AtoL}
\lambda_0 =\alpha_0\quad,\quad \lambda_i = (\alpha_i -1) \text{ for } i>1\ .
\end{equation}
With this choice, one obtains a formula that enables us to write the 
Wronskian index $\ell$ in terms of the exponents at the elliptic points.
\[
\ell = \frac{n(n-12)}{2} + 3 a_1 + 2(b_1+2b_2)\ ,
\]
where $a_1$ are the number of eigenvalues of $S$ equal to $-1$ and $b_j$ 
are the number of eigenvalues of $U$ equal to $e^{2\pi ij/3}$ (for 
$j=1,2$). The above formula, when na\"\i vely used by taking exponents 
from Table \ref{FExp} for $(6,0)$ theories gives $\ell=6$ -- this says 
that Eq. \eqref{AtoL} that relates the exponents of the RCFT to the 
$\lambda_i$ has to be modified. One of the $\lambda_i$ has to be shifted 
by one. This explains the observation in the rank six example (the 
tri-critical Ising model) studied in \cite{Govindarajan:2025rgh} that 
all quasi-characters obtained from the $(6,0)$ RCFT were of $(6,6)$ 
type. This appears to be a generic feature rather than something special 
to that example.

We restricted our attention to theories with Wronskian index $\ell<6$. 
Conjecturally, we believe all theories with $\ell>6$ should arise from 
theories with $\ell<6$. A large class of tenable solutions can be 
constructed using the theory of VVMFs. Given the rank $n$ VVMF 
associated with a tenable (or admissible) solution with Wronskian index 
$\ell<6$, we can construct a basis for VVMFs using the methods of 
\cite{Bantay:2007zz,Gannon:2013jua} applied to 
RCFTs\cite{Rayhaun:2023pgc,Govindarajan:2025rgh,Govindarajan:2025jlq}. 
Here by taking suitable linear combinations as explained in 
\cite{Govindarajan:2025jlq} for rank three VVMF, we can obtain 
admissible solutions of $(n,\ell+6p-6m$ as points in a $p$ dimensional 
polytope appearing on a co-dimension $m$. In work to appear, we have 
extended this work to the cases of $4$ and $5$ character 
theories\cite{Govindarajan:2026c}. While we lack a proof that this 
method exhausts all solutions with Wronskian index $>6$, given a parent 
$(n,\ell<6)$ theory, this method exhausts all possible admissible 
solutions in the class of multiplier associated with the parent theory. 
What is unproven is that there are theories with some unknown multiplier 
system and $\ell>6$) that do not have a $(n,\ell<6)$ theory with the 
same multiplier system.

\textbf{Remark:} The arXiv submission has a Mathematica package attached as an ancillary file.

\smallskip

\noindent\textbf{Acknowledgements} We would like to thank Aditya Jain, 
Abhiram Kidambi and Jagannath Santara for useful discussions and 
collaboration on earlier projects. Both of us would like to thank the 
organisers of the Chennai Strings Meeting 2025 held during March 2026 
for the opportunity to present this work.
\clearpage

\appendix

\section{Notation and Conventions}

\subsection{Notation}
\begin{center}
\begin{tabular}{c|l}
Symbol     &  Meaning \\ \midrule
  $\zeta_n$   &  $\exp(2\pi i/n)$ \\
  $c_n^m$  & $(\zeta_n)^m+ (\zeta_n)^{-m} = 2 \cos (2\pi m/m)$ \\
  $\mathcal{M}(p,q)$ & The $(p,q)$ Virasoro minimal model \\
  $G_{n,k}$ & The affine Kac-Moody algebra $G_n$ at level $k$  ($G$ is a semi-simple Lie algebra)\\
  $E_{7\frac12,1}$ & The IVOA that appears in the MMS $(2,0)$ list \\
  $E_{7\frac12,2}$ & The IVOA from \cite{Duan:2022ltz} \\
  $E_2(\tau)$, $E_4(\tau)$, $E_6(\tau)$ & The Eisenstein series \\
  $\Delta(\tau)$ & The modular discriminant\\
  $\ceff$ & $=(c-24 h_{\text{min}})$ for an RCFT \\
  $\heff$ &  $\left((h_0-h_{\text{min}}),\ldots, (h_{n-1}-h_{\text{min}})\right)$  \\
  $(n,\ell)$ & $n$ is the number of characters/rank of VVMF, $\ell$ is the Wronskian index \\ \bottomrule
\end{tabular}
\end{center}

\subsection{Modular Objects}

The modular forms  that we use in the paper are as given below.
\begin{align*}
E_2(\tau) &= 1 - 24\sum_{n=1}^{\infty} \frac{n\, q^n}{(1-q^n)} = 1-24q-72q^2-96q^3  +\cdots \\
E_4(\tau) &= 1 + 240\sum_{n=1}^{\infty} \frac{n^3\, q^n}{(1-q^n)} = 1 + 240 q + 2160 q^2 + 6720 q^3  + \cdots \\
E_6(\tau) &= 1 - 504\sum_{n=1}^{\infty} \frac{n^5\, q^n}{(1-q^n)} = 1 - 504 q - 16632 q^2 - 122976 q^3 +\cdots 
\end{align*}
\begin{align*}
\Delta(\tau) &= =  \frac{E_4^3 - E_6^2}{1728} = q - 24 q^2 + 252 q^3 - 1472 q^4 + 4830 q^5 + \cdots \\
J(\tau) &= \frac{E_4^3}{\Delta} = \frac{1}{q} + 744 + 196884 q + 21493760 q^2 +864299970 q^3+\cdots
\end{align*}
The Ramanujan-Serre covariant derivative, acting on modular forms of weight $w$ is,  
\[
D_w := q\frac{d}{dq} - \frac{w}{12} E_2(\tau)\ .
\]
In our examples, all our modular forms are typically weight zero and so we drop the subscript $w$ for ease of presentation
and the higher order derivatives are defined as
\[
D_w^{(n)} := \prod_{k=1}^n D_{w+2k-2}.
\]
\section{MLDEs in the $w$ coordinate}\label{MLDEw}

Define $w=\frac{1728}{J(\tau)}$ and let 
$\tilde{\rho}=\frac{\rho}{1728}$. We will give the MLDEs written in this 
coordinate along with values of non-accessory parameters that are 
determined in terms of the exponents. Let $\alpha_i$ (for 
$i=0,\ldots,(n-1)$ be the exponents about $w=0$ for an $n$-th order 
MLDE. Then, we define the symmetric polynomials, $e_j(\alpha)$, as 
follows:
\[
\prod_{i=0}^{n-1} (x-\alpha_i) = x^n + \sum_{j=1}^{n} (-1)^j\ 
e_j(\alpha)\ x^{n-j} \ .
\]

\noindent \textbf{n=4} 
The $(4,2)$ MLDE is given by
\begin{multline}
\left[\partial_w^4+\left(\frac{16/3}{w}+\frac{3}{w-1}\right)\partial^3_w+\left(\frac{185/36+\kappa}{w^2}+\frac{29/3-\kappa}{w(w-1)}+\frac{3/4}{(w-1)^2}\right)\partial^2_w+\right.  
\\
\left. 
\left(\frac{25/54+5\kappa/6+\mu}{w^3}+\frac{35/9-\kappa/3-\mu}{w^2(w-1)}+\frac{5/6+\kappa/2}{w(w-1)^2}\right)\partial_w+\frac{\nu}{w^4(w-1)^4} 
+\frac{\tilde{\rho}}{w^3(w-1)^2}\right]f(\tau)\, =0~.\\
\end{multline}

\noindent The $(4,4)$ MLDE is given by
\begin{multline}
\left[\partial_w^4+\left(\tfrac{17/3}{w}+\tfrac{3}{w-1}\right)\partial^3_w+\left(\tfrac{215/36+\kappa}{w^2}+\tfrac{61/6-\kappa}{w(w-1)}+\tfrac{3/4}{(w-1)^2}\right)\partial^2_w+\right. \\
\left. \left(\tfrac{35/54+5\kappa/6+\mu}{w^3}+\tfrac{40/9-\kappa/3-\mu}{w^2(w-1)}+\tfrac{5/6-\kappa/2}{w(w-1)^2}\right)\partial_w+\tfrac{\nu}{w^4(w-1)^2} +\tfrac{\tilde{\rho}}{w^3(w-1)^2}\right]f(\tau)\, =0~.
\end{multline}
Note that we have sent one of the accessory parameters to zero. The 
other parameters can be expressed in terms of symmetric polynomials of 
the exponents at $w=0$.
\[
\kappa(\ell)=e_2(\alpha)
-\frac{11-3\ell}{36},\ \nu= e_4(\alpha), \ 
\mu= e_3(\alpha)
+\frac{e_2(\alpha)}{6}
-\frac{5-\ell}{216}\ .
\]

\noindent $\textbf{n=5}$ 
The $(5,2)$ MLDE in the coordinate $w$ is
\begin{multline}
    \left[\partial_w^5+\left(\tfrac{26}{3w}+\tfrac{5}{w-1}\right)\partial^4_w+\left(\tfrac{635/36+\mu_2}{w^2}+\tfrac{181/6-\mu_2}{w(w-1)}+\tfrac{15}{4(w-1)^2}\right)\partial^3_w+ \right.\\ \left(\tfrac{805/108+5\mu_2/2+\mu_3}{w^3}\right. 
    \left.+ \tfrac{655/18-\mu_2-\mu_3}{w^2(w-1)}+\tfrac{51/4-3\mu_2/2}{w(w-1)^2}\right)\partial^2_w+\left(\tfrac{5/27+5\mu_2/9+5\mu_3/6+\mu_4}{w^4}\right.\\ 
    +\tfrac{55/9+10\mu_2/9-\mu_3/3-2\mu_4-\tilde{\rho}}{w^3(w-1)}
\left.\left.+\tfrac{50/9-5\mu_2/3-\mu_3/2+\mu_4+\tilde{\rho}}{w^2(w-1)^2}\right)\partial_w+\tfrac{\nu}{w^5(w-1)^2} \right]f(\tau)\, =0~.
\end{multline}
The non-accessory parameters can be expressed in terms of symmetric 
polynomials of the exponents at $w=0$.
\begin{multline}
        \mu_2=e_2(\alpha)-\tfrac{35-6\ell}{36}\ ,
\mu_3=-e_3(\alpha)+\tfrac{e_2(\alpha)}{2}
-\tfrac{55-7\ell}{216}\ , \\
 \mu_4=e_4(\alpha)-\tfrac{e_3(\alpha)}{6} +\tfrac{e_2(\alpha)}{36}
 -\tfrac{9-\ell}{1296}\ ,\  \nu=-e_5(\alpha)\ .
\end{multline}

\noindent $\textbf{n=6}$ 
The $(6,0)$ MLDE is given by
\begin{multline}
    \left[\partial_w^6+\left(\tfrac{25}{2w}+\tfrac{15/2}{w-1}\right)\partial^5_w+\left(\tfrac{1525/36+\mu_2}{w^2}+\tfrac{425/6-\mu_2}{w(w-1)}+\tfrac{45}{4(w-1)^2}\right)\partial^4_w+ \left(\tfrac{325/108+5\mu_2+\mu_3}{w^3}\right.\right. \\
    \left.+ \tfrac{1325/8-2\mu_2-\mu_3}{w^2(w-1)}+\tfrac{575/8-3\mu_2}{w(w-1)^2}+\tfrac{15/8}{(w-1)^3}\right)\partial^3_w+\left(\tfrac{2305/324+155\mu_2/36+5\mu_3/2+\mu_4}{w^4}\right. 
     \\ \left.  +\tfrac{9965/108+175\mu_2/36-\mu_3-2\mu_4}{w^3(w-1)} +\tfrac{25/4-3\mu_2/4}{w(w-1)^3}\right)\partial^2_w+
\left(\tfrac{655/10+55\mu_2/18+10\mu_3/9-7\mu_4/6-2\mu_5}{w^4(w-1)} \right.\\
\left.+\tfrac{2045/108-5\mu_2/2-5\mu_3/2-\mu_4/6+\mu_5}{w^3(w-1)^2}+\tfrac{95/36-5\mu_2/6+\mu_4/2}{w^2(w-1)^3}+\tfrac{5/324+5\mu_2/18+5\mu_3/9+5\mu_4/6+\mu_5}{w^5}\right)\partial_w\\\left. -\tfrac{\nu}{w^6(w-1)^3}-\tfrac{\rho}{w^5(w-1)^3} \right]f(\tau)\, =0~.
\end{multline}
The non-accessory parameters are expressed in terms of symmetric 
polynomials of the exponents at $w=0$
\begin{multline}\label{n6pars}
    \nu=e_6(\alpha)\ ,\ \mu_2=e_2(\alpha)-\tfrac{85-10\ell}{36}\ , \ 
\mu_3=-e_3(\alpha) + e_2(\alpha)-\tfrac{ 285-25\ell}{216},\\
 \mu_4=e_4(\alpha)-\tfrac{e_3(\alpha)}{2}+\tfrac{7e_2(\alpha)}{36}-\tfrac{194-15\ell}{1296}\ ,
\mu_5=-e_5(\alpha)+\tfrac{e_4(\alpha)}{6}-\tfrac{e_3(\alpha)}{36}+\tfrac{e_2(\alpha)}{216}-\tfrac{14-\ell}{7776}
\end{multline}

\section{List of allowed exponents for $n=6$ MLDEs}

\subsection{The $(6,0)$ list}
\begin{longtable}{|c|>{\raggedright\arraybackslash}p{0.95\linewidth}|}
\toprule
$N$& $\alpha_i \text{ (mod } 1)$\\
\midrule
\endfirsthead
\midrule
6 & $(0, \frac{1}{6}, \frac{1}{3}, \frac{1}{2}, \frac{2}{3}, \frac{5}{6})$ \\
\midrule
8 & $(0, \frac{1}{8}, \frac{1}{4}, \frac{1}{2}, \frac{3}{4}, \frac{7}{8})$, $(0, \frac{1}{8}, \frac{3}{8}, \frac{1}{2}, \frac{5}{8}, \frac{7}{8})$, $(0, \frac{1}{4}, \frac{3}{8}, \frac{1}{2}, \frac{5}{8}, \frac{3}{4})$ \\
\midrule
10 & $(0, \frac{1}{10}, \frac{2}{5}, \frac{1}{2}, \frac{3}{5}, \frac{9}{10})$, $(0, \frac{1}{5}, \frac{3}{10}, \frac{1}{2}, \frac{7}{10}, \frac{4}{5})$ \\
\midrule
12 & $(0, \frac{1}{12}, \frac{1}{3}, \frac{1}{2}, \frac{3}{4}, \frac{5}{6})$, $(0, \frac{1}{6}, \frac{1}{4}, \frac{1}{2}, \frac{2}{3}, \frac{11}{12})$, $(0, \frac{1}{6}, \frac{5}{12}, \frac{1}{2}, \frac{2}{3}, \frac{3}{4})$, $(0, \frac{1}{4}, \frac{1}{3}, \frac{1}{2}, \frac{7}{12}, \frac{5}{6})$, $(\frac{1}{12}, \frac{1}{6}, \frac{1}{3}, \frac{5}{12}, \frac{2}{3}, \frac{5}{6})$, $(\frac{1}{6}, \frac{1}{3}, \frac{7}{12}, \frac{2}{3}, \frac{5}{6}, \frac{11}{12})$ \\
\midrule
14 & $(\frac{1}{14}, \frac{1}{7}, \frac{2}{7}, \frac{4}{7}, \frac{9}{14}, \frac{11}{14})$, $(\frac{3}{14}, \frac{5}{14}, \frac{3}{7}, \frac{5}{7}, \frac{6}{7}, \frac{13}{14})$ \\
\midrule
16 & $(0, \frac{1}{16}, \frac{5}{16}, \frac{9}{16}, \frac{3}{4}, \frac{13}{16})$, $(0, \frac{1}{16}, \frac{7}{16}, \frac{1}{2}, \frac{9}{16}, \frac{15}{16})$, $(0, \frac{3}{16}, \frac{1}{4}, \frac{7}{16}, \frac{11}{16}, \frac{15}{16})$, $(0, \frac{3}{16}, \frac{5}{16}, \frac{1}{2}, \frac{11}{16}, \frac{13}{16})$, $(\frac{1}{16}, \frac{1}{8}, \frac{5}{16}, \frac{9}{16}, \frac{5}{8}, \frac{13}{16})$, $(\frac{1}{16}, \frac{3}{16}, \frac{1}{4}, \frac{9}{16}, \frac{11}{16}, \frac{3}{4})$, $(\frac{1}{16}, \frac{1}{4}, \frac{5}{16}, \frac{1}{2}, \frac{9}{16}, \frac{13}{16})$, $(\frac{3}{16}, \frac{3}{8}, \frac{7}{16}, \frac{11}{16}, \frac{7}{8}, \frac{15}{16})$, $(\frac{3}{16}, \frac{7}{16}, \frac{1}{2}, \frac{11}{16}, \frac{3}{4}, \frac{15}{16})$, $(\frac{1}{4}, \frac{5}{16}, \frac{7}{16}, \frac{3}{4}, \frac{13}{16}, \frac{15}{16})$ \\
\end{longtable}

\subsection{The $(6,3)$ list}

Surprisingly, this is a lot smaller than the $(6,0)$ list.
\begin{longtable}{|c|l|}
\toprule
$N$& $\alpha_i \text{ (mod } 1)$\\
\midrule
\endfirsthead
\midrule
7 & $(\frac{1}{7},\frac{2}{7},\frac{3}{7},\frac{4}{7},\frac{5}{7},\frac{6}{7})$ \\
\midrule
9 & $(\frac{1}{9},\frac{2}{9},\frac{4}{9},\frac{5}{9},\frac{7}{9},\frac{8}{9})$\\
\midrule
14 & $(\frac{1}{14},\frac{3}{14},\frac{5}{14},\frac{9}{14},\frac{11}{14},\frac{13}{14})$ \\
\midrule
18 & $(\frac{1}{18},\frac{5}{18},\frac{7}{18},\frac{11}{18},\frac{13}{18},\frac{17}{18})$ \\
\midrule
21 & $(\frac{1}{21},\frac{4}{21},\frac{10}{21},\frac{13}{21},\frac{16}{21},\frac{19}{21})$, $(\frac{2}{21},\frac{5}{21},\frac{8}{21},\frac{11}{21},\frac{17}{21},\frac{20}{21})$ \\
\midrule
42 & $(\frac{1}{42},\frac{13}{42},\frac{19}{42},\frac{25}{42},\frac{31}{42},\frac{37}{42})$, $(\frac{5}{42},\frac{11}{42},\frac{17}{42},\frac{23}{42},\frac{29}{42},\frac{41}{42})$ \\
\bottomrule
\caption{List of (6,3) exponents (mod 1)}
\label{tab:KLP63}
\end{longtable}
\section{Lists of admissible solutions}

Below, we list all admissible solutions with $c_{\text{eff}}\leq 24$ that we obtain from MLDEs with one accessory parameter. We organise the lists in multiple ways. The serial number is given as $\bar{N}.n$ where $\bar{N}$ is the order of the normalized $T$-matrix defined by $\bar{T}=T/T_{00}$. Further, $n$ is the serial number associated with the admissible solution. The second column gives $\ceff$ and the third column gives the effective weights, $\heff$.  It is not obvious whether an admissible solution comes from a unitary or a non-unitary RCFT. We are able to fix this once we obtain the $S$-matrix that is given in column 6. Column 4 gives the value of the accessory parameter and Column 5 gives the value of $m_1$ as well as the degeneracy. The degeneracy is given as a product $\mathbf{y_i}\cdot a_{i,0}$ where $a_{i,0}$ is the minimal value needed for admissibility and $\mathbf{y}_i$ is an additional factor that is fixed by requiring the multiplicity to be square-free and/or acceptable fusion rules. In Column 5, when we find a tenable solutions that is potentially non-unitary RCFT, we identify the identity operator by adding to star to the degeneracy (which has to be one). Sometimes, there is more than one choice and then we show all of them. See entry 7.2 in the list of $(4,2)$ admissible solutions.

The admissible solutions are classified into three groups through the color coding in Column 5. Green indicates a tenable solution, Red indicates a solution that has degenerate vacuum (also called IVOA type\cite{Chandra:2018pjq}) and an empty entry indicates that the admissible solution has fusion rules that are untenable. Finally, in Column 6 we provide additional information that we could obtain. In some cases, we identify tenable solutions with known RCFTs. In others, we identify the MTC class of the tenable solution in the notation of \cite{Ng:2023}

\subsection{\textbf{(4,2)} admissible solutions}
\begin{longtable}{|c ||c |c ||c |l |c ||c|}
\toprule
S. No. & $c_{\text{eff}}$ & $H_{\text{eff}}$ & $\rho$ & $(\mathfrak{m},\mathbf{y_i}\cdot a_{i,0})$ &   $S$-matrix& MTC/RCFT\\
\midrule
\endfirsthead
4.1 & $5$ & $\frac{1}{4}, \frac{1}{2}, \frac{3}{4}$ & $-\frac{37}{3}$ & $(31, \mathbf{2}\cdot1, \mathbf{8}\cdot1, \mathbf{16}\cdot1)$&{\color{mgreen}$[4,1]$}& $D_{4,1}\otimes A_{1,1}$\\
4.2 & $11$ & $\frac{1}{4}, \frac{1}{2}, \frac{7}{4}$ & $\frac{395}{3}$ & $(273, 7, 5, 15)$ & & \\
4.3 & $11$ & $\frac{3}{4}, \frac{1}{2}, \frac{5}{4}$& $-\frac{109}{3}$ & $(161, \mathbf{8}\cdot7, \mathbf{8}\cdot1, \mathbf{64}\cdot7)$& {\color{mgreen}$[4,1]$}& $D_{4,1}\otimes  E_{7,1}$\\

5.1 & $\frac{16}{5}$ & $\frac{1}{5}, \frac{2}{5}, \frac{3}{5}$ & $-\frac{104}{15}$ & $(15, 1^*, 7, 7)$& {\color{mgreen}$[5,1]$}& $G_{2,1}\otimes \mathcal{M}(5,2)$\\
 5.2& $\frac{28}{5}$ & $\frac{1}{5}, \frac{3}{5}, \frac{4}{5}$ & $-\frac{248}{15}$ & $(53, 1^*, 26, 26)$& {\color{mgreen}$[5,1]$}&$F_{4,1}\otimes \mathcal{M}(5,2)$ \\
5.3& $\frac{34}{5}$ & $\frac{2}{5}, \frac{3}{5}, \frac{4}{5}$ & $-\frac{1708}{75}$ & $(108, 2, \mathbf{5}\cdot2, \mathbf{5}\cdot7)$ & {\color{red}$[5,2]$}& \\
5.4& $\frac{46}{5}$ & $\frac{1}{5}, \frac{2}{5}, \frac{8}{5}$ & $\frac{8228}{75}$ & $(254, 16, 11, 242)$ & & \\
5.5& $\frac{46}{5}$ & $\frac{3}{5}, \frac{2}{5}, \frac{6}{5}$ & $-\frac{988}{75}$ & $(206, 16, \mathbf{5}\cdot1, \mathbf{5}\cdot52)$ & {\color{red}$[5,2]$}& \\
\end{longtable}

\subsection{\textbf{(4,4)} admissible solutions}
\begin{longtable}{|c ||c |c ||c |l|c ||c|}
\toprule
S. No. & $\ceff$ & $\heff$ & $\rho$ & $(\mathfrak{m},a_{i,0})$ & $S$& MTC\\
\midrule
\endfirsthead
4.1 & $13$ & $\frac{5}{4}, \frac{1}{2}, \frac{3}{4}$ & $-\frac{388}{3}$ & $(199, \mathbf{32}\cdot21, \mathbf{4}\cdot1, \mathbf{8}\cdot5)$& {\color{red}[4,1]}& \\
4.2 & $19$ & $\frac{3}{4},\frac{3}{2},\frac{5}{4}$& $-\frac{1108}{3}$ & $(121, \mathbf{8}\cdot1, \mathbf{512}\cdot7,\mathbf{256}\cdot3 )$& {\color{red}[4,1]}& \\
5.1 & $\frac{38}{5}$ & $\frac{4}{5},\frac{1}{5}, \frac{3}{5}$& $-\frac{752}{15}$ & $(55, 78, 1, 12)$& {\color{mgreen}[5,3]}& $8_{\frac{2}{5},47.36}^{5,148}$\\
5.2 & $\frac{56}{5}$ & $\frac{6}{5}, \frac{2}{5}, \frac{3}{5}$ & $-\frac{5776}{75}$ & $(228, \mathbf{5}\cdot66, \mathbf{5}\cdot1, 22)$& {\color{red}$[5,1]$}& \\
5.3 & $\frac{68}{5}$ & $\frac{6}{5}, \frac{3}{5}, \frac{4}{5}$ & $-\frac{11824}{75}$ & $(256, \mathbf{5}\cdot88,\mathbf{5}\cdot 4, 66)$& {\color{red}$[5,1]$}& \\
5.4 & $\frac{74}{5}$ & $\frac{2}{5}, \frac{3}{5}, \frac{9}{5}$ & $\frac{3728}{75}$ & $(411, 7, \mathbf{25}\cdot1,\mathbf{25}\cdot 357)$& {\color{red}$[5,2]$}& \\
5.5 & $\frac{74}{5}$ & $\frac{7}{5}, \frac{3}{5}, \frac{4}{5}$ & $-\frac{496}{3}$ & $(334, 684, 4, 49)$ & {\color{red}$[5,2]$}& \\
5.6 & $\frac{86}{5}$ & $\frac{3}{5}, \frac{2}{5}, \frac{11}{5}$& $\frac{23888}{75}$ & $(574, 26, 1, 7774)$& & \\
\end{longtable}

\subsection{\textbf{4 character $S$-matrices}}
\textbf{}Indices giving positive integral fusion coefficients are denoted as $ind_p$, 

\begin{enumerate}[wide, labelwidth=!, labelindent=-3pt]
    \item \[
                 \bar{S}[4,1]=
\left(%
\right),\qquad
S_{00}=0.37175, \qquad
 m_i=(1,1,1,1),\qquad
ind_p=0,2.
\end{align*}

\end{enumerate}

\subsection{\textbf{(5,2)} admissible solutions}

Some of the theories are obtained as Hecke transforms of known minimal models and appear in the work of \cite{Duan:2022ltz}.

\begin{longtable}{|c ||c |c ||c |l |c ||c|}
\toprule
S. No. & $\ceff$ & $\heff$ & $\rho$ & $(\mathfrak{m},a_{i,0})$ & $S$& MTC/RCFT\\
\midrule
\endfirsthead
5.1 & $\frac{16}{5}$ & $\frac{1}{5}, \frac{2}{5}, \frac{3}{5}, \frac{4}{5}$ & $-\frac{104}{5}$ & $(0, 2, 1, 8, \mathbf{5}\cdot1)$& \color{mgreen}{[5,1]}& $T_{2}(\mathcal{M}(5,2)^{\otimes4})$ \\
5.2 & $\frac{64}{5}$ & $\frac{1}{5}, \frac{2}{5}, \frac{8}{5}, \frac{9}{5}$ & $-\frac{2248}{25}$ & $(344, 4, 14, 11, 28)$ & & \\
5.3 & $\frac{64}{5}$ & $\frac{1}{5}, \frac{3}{5}, \frac{7}{5}, \frac{9}{5}$ & $-\frac{2536}{25}$ & $(302, 9, 7, 2808, 637)$& & \\
5.4 & $\frac{64}{5}$ & $\frac{4}{5},\frac{8}{5},\frac{2}{5},\frac{6}{5}$& $-\frac{3112}{25}$ & $64,\mathbf{25}\cdot2,\mathbf{25}\cdot33,4,\mathbf{25}\cdot12)$& \color{mgreen}{[5,1]}& \\
5.5 & $\frac{64}{5}$ & $\frac{4}{5}, \frac{3}{5}, \frac{7}{5}, \frac{6}{5}$& $-136$ & $(32, 54,8,1728,\mathbf{5}\cdot56)$& \color{mgreen}{[5,1]}& $T_{8}(\mathcal{M}(5,2)^{\otimes4})$\\
5.6 & $\frac{88}{5}$ & $\frac{1}{5},\frac{8}{5},\frac{14}{5},\frac{2}{5}$& $\frac{8984}{25}$ & $(506,11,2662,58564,121)$& & \\
5.7 & $\frac{88}{5}$ & $\frac{11}{5},\frac{8}{5},\frac{4}{5},\frac{2}{5}$& $-\frac{3112}{25}$ & $(660, 476, 22, 33, 231)$ & & \\
5.8 & $\frac{88}{5}$ & $\frac{6}{5},\frac{8}{5},\frac{4}{5},\frac{7}{5}$& $-\frac{6568}{25}$ & $(220,\mathbf{25}\cdot22,\mathbf{25}\cdot77,11,\mathbf{25}\cdot44)$& \color{red}{[5,2]}& $T_{11}(\mathcal{M}(5,2)^{\otimes4})$ \\
\end{longtable}

\subsection{$5$ character $S$-matrices}
Indices giving positive integral fusion coefficients are denoted as $ind_p$.

\begin{enumerate}[wide, labelwidth=!, labelindent=-3pt]
    \item \begin{align*}
                 \bar{S}[5,1]&=
\left(%
\right)
 ,\\
S_{00}&=\frac12(-2+2c_{11}^1-4c_{11}^2+c_{11}^3),\qquad m_i=(1,1,1,1,1),\qquad
ind_p=1.
\end{align*}
\end{enumerate}

\subsection{\textbf{(6,0)} admissible solutions}

In the last column, we have partial identifications for our tenable solutions, either in the form of an MTC label, or an RCFT\footnote{We thank Kaiwen Sun for identifying a few of the RCFTs.} . Labels such as $A_{1,20}^D$ with a superscript $D$ refer to the off-diagonal D-type invariant. A superscript such as $A_{11,1}^X$ refers to a sub-theory obtained by combining characters whose weights differ by integers. The $U(1)_k$ RCFT refers to the $U(1)$ theory with $c=1$, $h=q^2/4k$ with $q=0,\pm1,\ldots,\pm k$.

\begin{longtable}{|c ||c |c ||c |>{\raggedright\arraybackslash}p{0.4\linewidth}|  c ||l|}
\toprule
S. No. & $\ceff$ & $\heff$ & $\rho$ & $(\mathfrak{m},a_{i,0})$ & $S$   & MTC/RCFT\\
\midrule
\endfirsthead
$11.1$ & $\frac{10}{11}$ & $\frac{1}{11},\frac{2}{11},\frac{5}{11},\frac{7}{11},\frac{15}{11}$& $\frac{1761725}{8696754}$ & $(1,1,1,1,1,1^*)$&   {\color{mgreen}[11,1]}& $\mathcal{M}(22,3)^D$\\
$11.2$ & $\frac{30}{11}$ & $\frac{1}{11},\frac{3}{11},\frac{6}{11},\frac{10}{11},\frac{15}{11}$& $-\frac{19425}{322102}$ & $(3,3,5,7,9,\mathbf{11}\cdot1)$&  {\color{mgreen}[11,2]}& $7_{\frac{30}{11},135.7}^{11,157}$\\
&&&&&& $(A_{1,20}^D)$ \\
$11.3$ & $10$ & $\frac{12}{11},\frac{14}{11},\frac{15}{11},\frac{5}{11},\frac{9}{11}$ & $\frac{21357125}{8696754}$ & $(120,\mathbf{11}\cdot15,\mathbf{11}\cdot30,\mathbf{11}\cdot42,\mathbf{11}\cdot1,\mathbf{11}\cdot5)$&   {\color{mgreen}[11,3]}& $11_{2,11.}^{11,568}$\\
 &  &  &  && &  $(A_{10,1})$ \\
$13.1$ & $\frac{10}{13}$ & $\frac{1}{13},\frac{3}{13},\frac{6}{13},\frac{10}{13},\frac{15}{13}$& $\frac{24339175}{260647686}$ & $(1,1,1,1,1,1^*)$&   {\color{mgreen}[13,1]}& $\mathcal{M}(13,2)$\\
$13.2$ & $\frac{90}{13}$ & $\frac{14}{13},\frac{15}{13},\frac{5}{13},\frac{9}{13},\frac{12}{13}$ & $\frac{32588325}{9653618}$ & $(45,90,42,\mathbf{3}\cdot3,35,75)$&   {\color{red}[13,2]}& $osp(1|10)_1$\\
$13.3$ & $\frac{190}{13}$ & $\frac{18}{13},\frac{19}{13},\frac{21}{13},\frac{10}{13},\frac{12}{13}$ & $\frac{3824637775}{260647686}$ & $(190,\mathbf{5}\cdot209,\mathbf{5}\cdot528,\mathbf{19}\cdot80,\mathbf{19}\cdot3,\mathbf{19}\cdot10)$&   {\color{red}[13,3]}& $E_{7\frac{1}{2},2}$\\

$15.1$ & $14$ & $\frac{6}{5},\frac{7}{15},\frac{5}{3},\frac{9}{5},\frac{13}{15}$ & $-\frac{484393}{33750}$ & $(224,\mathbf{5}\cdot91,\mathbf{15}\cdot1,\mathbf{3}\cdot1001,\mathbf{65}\cdot77,\mathbf{15}\cdot7)$&   {\color{mgreen}[15,1]}& $A_{14,1}^X$\\
$15.2$ & $22$ & $\frac{6}{5},\frac{22}{15},\frac{5}{3},\frac{9}{5},\frac{28}{15}$ & $\frac{711293}{11250}$ & $(22,\mathbf{25}\cdot11,\mathbf{4050}\cdot1,\mathbf{81}\cdot88,\mathbf{275}\cdot56,\mathbf{4050}\cdot11)$&   {\color{mgreen}[15,1]}& \\

$20.1$ & $1$ & $\frac{1}{20},\frac{1}{5},\frac{5}{4},\frac{9}{20},\frac{4}{5}$ & $\frac{22939}{540000}$ & $(1,1,1,\mathbf{2}\cdot1,1,1)$&   {\color{mgreen}[20,1]}& $10_{1,10.}^{20,667}$\\
&&&&&& $(U(1)_5)$ \\
$20.2$ & $9$ & $\frac{21}{20},\frac{6}{5},\frac{5}{4},\frac{9}{20},\frac{4}{5}$ & $\frac{81081}{20000}$ & $(99,\mathbf{10}\cdot12,\mathbf{5}\cdot42,\mathbf{12}\cdot21,\mathbf{10}\cdot1,\mathbf{5}\cdot9)$&   {\color{mgreen}[20,1]}& $10_{1,10.}^{20,667}$\\
 &  &  &  && & $(A_{9,1})$ \\
$21.1$ & $\frac{6}{7}$ & $\frac{1}{21},\frac{1}{7},\frac{4}{3},\frac{10}{21},\frac{5}{7}$ & $\frac{119557}{2117682}$ & $(0,1,1,1,1,1)$ &   {\color{mgreen}[21,1]}& $9_{\frac{6}{7},27.88}^{21,155}$\\
&&&&&& $(\mathcal{M}(7,6)^D)$ \\
$21.2$ & $\frac{78}{7}$ & $\frac{13}{21},\frac{6}{7},\frac{4}{3},\frac{25}{21},\frac{9}{7}$& $\frac{21513349}{2117682}$ & $(78,\mathbf{27}\cdot1,78,\mathbf{351}\cdot1,\mathbf{27}\cdot13,\mathbf{13}\cdot50)$&   {\color{mgreen}[21,1]}& $9_{\frac{22}{7},27.88}^{21,204}$\\
&&&&&& $(E_{6,2})$\\
$24.1$ & $1$ & $\frac{1}{24},\frac{1}{6},\frac{3}{8},\frac{3}{2},\frac{2}{3}$ & $\frac{1925}{9216}$ & $(1,1,1,1,2,1)$ &   {\color{mgreen}[24,1]}& $U(1)_6^X$\\
$24.2$ & $9$ & $\frac{1}{24},\frac{7}{6},\frac{3}{8},\frac{1}{2},\frac{8}{3}$ & $\frac{226765}{9216}$ & $(309,1,13,91,21,57)$&   & \\
$24.3$ & $11$ & $\frac{11}{24},\frac{5}{6},\frac{9}{8},\frac{3}{2},\frac{4}{3}$& $\frac{1925}{9216}$ & $(143,\mathbf{12}\cdot1,\mathbf{6}\cdot11,\mathbf{4}\cdot55,\mathbf{44}\cdot21,\mathbf{3}\cdot165)$&    {\color{mgreen}[24,1]}& $A_{11,1}^X$\\

$28.1$ & $\frac{5}{7}$ & $\frac{1}{28},\frac{5}{28},\frac{2}{7},\frac{10}{7},\frac{3}{4}$ & $\frac{8155675}{101648736}$ & $(1,1,1^*,1,1^*,1)$&   {\color{mgreen}[28,1]}& \\
$28.2$ & $\frac{15}{7}$ & $\frac{3}{28},\frac{5}{4},\frac{2}{7},\frac{15}{28},\frac{6}{7}$ & $\frac{568425}{3764768}$ & $(3,\mathbf{2}\cdot1,\mathbf{2}\cdot3,3,\mathbf{2}\cdot2,5)$& {\color{mgreen}[28,2]}&  $6_{\frac{15}{7},18.59}^{28,289}$\\
 &  &  &  && &  $(A_{1,5})$\\
$28.3$ & $\frac{55}{7}$ & $\frac{11}{28},\frac{5}{4},\frac{5}{7},\frac{27}{28},\frac{8}{7}$& $\frac{285817675}{101648736}$ & $(55,\mathbf{2}\cdot5,\mathbf{2}\cdot66,\mathbf{11}\cdot4,\mathbf{2}\cdot55,165)$&   {\color{mgreen}[28,2]}& $6_{\frac{55}{7},18.59}^{28,108}$ \\
&&&&&& $(C_{5,1})$\\
$32.1$ & $\frac{3}{4}$ & $\frac{1}{32},\frac{5}{32},\frac{1}{4},\frac{1}{2},\frac{7}{4}$ & $\frac{2258025}{4194304}$ & $(1,1,1,1,1,1)$ &   & \\
$32.2$ & $1$ & $\frac{1}{32},\frac{1}{8},\frac{9}{32},\frac{49}{32},\frac{25}{32}$ & $\frac{8716763}{113246208}$ & $(1,1,1,1,1,1)$ & & \\
$32.3$ & $\frac{9}{4}$ & $\frac{3}{32},\frac{1}{4},\frac{15}{32},\frac{3}{2},\frac{3}{4}$ & $\frac{2027025}{4194304}$ & $(3,1,3,\mathbf{4}\cdot1,7,5)$&   {\color{mgreen}[32,1]}& $7_{\frac{9}{4},27.31}^{32,918}$\\
&&&&&& $(A_{1,6}^X)$ \\
$32.4$ & $\frac{39}{4}$ & $\frac{13}{32},\frac{3}{4},\frac{33}{32},\frac{3}{2},\frac{5}{4}$& $\frac{2027025}{4194304}$ & $(78,\mathbf{2}\cdot3,\mathbf{13}\cdot5,\mathbf{16}\cdot13,429,\mathbf{11}\cdot39)$&   {\color{mgreen}[32,1]}& $7_{\frac{7}{4},27.31}^{32,912}$ \\
&&&&&& $(C_{6,1}^X)$ \\
$32.5$ & $15$ & $\frac{39}{32},\frac{15}{32},\frac{55}{32},\frac{7}{8},\frac{63}{32}$ & $-\frac{80334975}{4194304}$ & $(255,35,1,273,15,715)$ &   & \\
$32.6$ & $\frac{85}{4}$ & $\frac{5}{4},\frac{3}{2},\frac{51}{32},\frac{55}{32},\frac{7}{4}$& $\frac{7986271475}{113246208}$ & $(85,357,\mathbf{17}\cdot275,\mathbf{256}\cdot35,\mathbf{256}\cdot51,\mathbf{17}\cdot495)$ &  {\color{red}[32,2]}& \\

$35.1$ & $\frac{782}{35}$ & $\frac{9}{7},\frac{16}{35},\frac{13}{5},\frac{20}{7},\frac{31}{35}$& {\tiny $ -\tfrac{1215377456129}{3970653750}$} & $(595,1496,17,\mathbf{34}\cdot19778,\mathbf{62}\cdot32538,187)$&   {\color{mgreen}[35,1]}& $6_{\frac{222}{35},33.63}^{35,224}$ \\
$35.2$ & $\frac{782}{35}$ & $\frac{9}{7},\frac{51}{35},\frac{8}{5},\frac{13}{7},\frac{66}{35}$&{\tiny $\frac{294432149791}{3970653750}$} & $(0,782,\mathbf{46}\cdot78,\mathbf{17}\cdot299,\mathbf{23}\cdot1122,60996)$&   {\color{mgreen}[35,1]}& $6_{\frac{222}{35},33.63}^{35,224}$ \\
$40.1$ & $\frac{4}{5}$ & $\frac{1}{40},\frac{1}{8},\frac{2}{5},\frac{21}{40},\frac{13}{8}$ & $\frac{3590543}{17280000}$ & $(0,1,1,1,1,1)$ &   & \\
$40.2$ & $\frac{12}{5}$ & $\frac{3}{40},\frac{1}{5},\frac{3}{8},\frac{63}{40},\frac{7}{8}$ & $\frac{75669}{640000}$ & $(3,1,3,1,2,3)$ &  &\\
$40.3$ & $\frac{68}{5}$ & $\frac{9}{8},\frac{17}{40},\frac{13}{8},\frac{4}{5},\frac{77}{40}$ & $-\frac{235751257}{17280000}$ & $(136,17,1,14,119,13)$ &   & \\
$40.4$ & $\frac{91}{5}$ & $\frac{11}{10},\frac{49}{40},\frac{3}{2},\frac{8}{5},\frac{13}{8}$ & $\frac{784978103}{17280000}$ & $(91,\mathbf{52}\cdot7,\mathbf{64}\cdot13,\mathbf{4}\cdot637,7007,\mathbf{64}\cdot77)$&   {\color{mgreen}[40,1]}& $8_{\frac{11}{5},14.47}^{40,824}$\\

$48.1$ & $1$ & $\frac{1}{48},\frac{3}{16},\frac{1}{3},\frac{25}{48},\frac{27}{16}$ & $\frac{905527}{1769472}$ & $(1,1,1,1,1,1)$ &  &\\
$48.2$ & $\frac{15}{2}$ & $\frac{1}{48},\frac{1}{3},\frac{5}{2},\frac{11}{16},\frac{5}{6}$ & $\frac{28773325}{589824}$ & $(129,11,7,99,117,319)$ &   & \\
$48.3$ & $\frac{39}{2}$ &$\frac{7}{6},\frac{65}{48},\frac{3}{2},\frac{5}{3},\frac{27}{16}$& $\frac{32965933}{589824}$ & $(78,\mathbf{27}\cdot13,\mathbf{1728}\cdot1,\mathbf{3}\cdot975,\mathbf{27}\cdot273,\mathbf{128}\cdot91)$&  {\color{mgreen}[48,1]}& $9_{\frac{7}{2},12.}^{48,342}$\\
$48.4$ & $23$ & $\frac{21}{16},\frac{23}{48},\frac{5}{3},\frac{45}{16},\frac{95}{48}$ & $-\frac{297613169}{1769472}$ & $(575,253,1,3542,7106,77)$ &   & \\

$80.1$ & $\frac{7}{10}$ & $\frac{3}{80},\frac{1}{10},\frac{7}{16},\frac{3}{2},\frac{3}{5}$ & $\frac{115305407}{1105920000}$ & $(0,1,1,1,1,1)$ &   {\color{mgreen}[80,1]}&  $6_{\frac{7}{10},14.47}^{80,111}$\\
 & & & & & &($\mathcal{M}(5,4)$)\\
$80.2$ & $\frac{69}{10}$ & $\frac{1}{80},\frac{11}{5},\frac{1}{2},\frac{7}{10},\frac{13}{16}$ & $\frac{716296581}{40960000}$ & $(106,1,78,7,34,119)$ &   & \\
$80.3$ & $\frac{133}{10}$ & $\frac{57}{80},\frac{9}{10},\frac{21}{16},\frac{3}{2},\frac{7}{5}$& {\tiny $\frac{14299561367}{1105920000}$} & $(133,\mathbf{8}\cdot6,81,\mathbf{152}\cdot77,\mathbf{19}\cdot7,\mathbf{19}\cdot133)$ &   {\color{mgreen}[80,1]}& $6_{\frac{53}{10},14.47}^{80,884}$\\
 & & & & & &$(E_{7,2})$\\
 \bottomrule
 \caption{The list of $(6,0)$ admissible solutions }\label{tab:60}
\end{longtable}
\subsection{\textbf{6 character $S$-matrices}}
\begin{enumerate}[wide, labelwidth=!, labelindent=-3pt]
    \item
    \begin{align*}
    \bar{S}[11,1]&=
\left(
\begin{smallmatrix}
 1 & 2 c^2_{11}-c^4_{11}-1 & \gamma_1 & -2c^3_{11}-2 c^4_{11}-2 &\gamma_1+c^3_{11} & 1-c^2_{11} \\
 c^2_{11}-c^4_{11}-1 & c^2_{11}-1 &\gamma_1+c^3_{11} & 2c^3_{11}+2 c^4_{11}+2 & -1 & -\gamma_1\\
 \gamma_1 &\gamma_1+c^3_{11} & -c^2_{11}+c^4_{11}+1 & 2c^3_{11}+2 c^4_{11}+2 & 1-c^2_{11} & 1 \\
 -c^3_{11}-c^4_{11}-1 & c^3_{11}+c^4_{11}+1 & c^3_{11}+c^4_{11}+1 & -c^3_{11}-c^4_{11}-1 & -c^3_{11}-c^4_{11}-1 & c^3_{11}+c^4_{11}+1 \\
 \gamma_1+c^3_{11} & -1 & 1-c^2_{11} & -2c^3_{11}-2 c^4_{11}-2 & -\gamma_1& c^2_{11}-c^4_{11}-1 \\
 1-c^2_{11} & -\gamma_1& 1 & 2c^3_{11}+2 c^4_{11}+2 & c^2_{11}-c^4_{11}-1 &\gamma_1+c^3_{11} \\
\end{smallmatrix}
\right)\\
 \gamma_1&= c^1_{11}+c^3_{11}+2 c^4_{11}+2\\
S_{00}&=0.50729, \qquad
 m_i=(1,1,1,2,1,1),\qquad
ind_p=5.
\end{align*}
\item    
\begin{align*}
    \bar{S}[11,2]&=\left(
\begin{smallmatrix}
 1 & \gamma_1 &\gamma_1+c^1_{11} &\gamma_2 &\gamma_2+c^2_{11} &2\gamma_3 \\
 \gamma_1 &\gamma_2+c^2_{11} &\gamma_2 & 1 &-\gamma_1 -c^1_{11}& -2\gamma_3 \\
\gamma_1+c^1_{11} &\gamma_2 & -\gamma_1& -\gamma_2-c^2_{11} & 1 &2\gamma_3\\
\gamma_2 & 1 & -\gamma_2-c^2_{11} &\gamma_1+c^1_{11} & \gamma_1 & -2\gamma_3\\
\gamma_2+c^2_{11} &-\gamma_1 -c^1_{11}& 1 & \gamma_1 &-\gamma_2 &2\gamma_3\\
\gamma_3 & -\gamma_3 &\gamma_3 & -\gamma_3 &\gamma_3 & -\gamma_3 \\
\end{smallmatrix}
\right)\\
 \gamma_1&=1-c^5_{11},\qquad \gamma_2=2 c^1_{11}+c^2_{11}+c^3_{11}+2 , \qquad \gamma_3= c^1_{11}+c^2_{11}+1,\\
S_{00}&=0.08582, \qquad
 m_i=(1,1,1,1,1,2),\qquad
ind_p=0.
\end{align*}
\item \begin{align*}
    \bar{S}[11,3]&=
\left(
\begin{smallmatrix}
  1 & 2 & 2 & 2 & 2 & 2 \\
 1 & c^2_{11} & c^1_{11} & c^4_{11} & c^3_{11} & c^5_{11} \\
 1 & c^1_{11} & c^5_{11} & c^2_{11} & c^4_{11} & c^3_{11} \\
 1 & c^4_{11} & c^2_{11} & c^3_{11} & c^5_{11} & c^1_{11} \\
 1 & c^3_{11} & c^4_{11} & c^5_{11} & c^1_{11} & c^2_{11} \\
 1 & c^5_{11} & c^3_{11} & c^1_{11} & c^2_{11} & c^4_{11} \\
\end{smallmatrix}
\right)\\
S_{00}&=0.30151, \qquad
 m_i=(1,2,2,2,2,2),\qquad
ind_p=0.
\end{align*}
\item \begin{align*}
    \bar{S}[13,1]&=
\left(
\begin{smallmatrix}
  1 &\gamma_1 & -\gamma_1+ c^1_{13} & c^2_{13}+c^3_{13}+c^4_{13} & -c^3_{13}-c^4_{13} & c^3_{13} \\
\gamma_1 & -c^3_{13}-c^4_{13} & -c^3_{13} &\gamma_1-c^1_{13} & -1 & -c^2_{13}-c^3_{13}-c^4_{13} \\
-\gamma_1+ c^1_{13}& -c^3_{13} & -1 & c^3_{13}+c^4_{13} & c^2_{13}+c^3_{13}+c^4_{13} &\gamma_1 \\
 c^2_{13}+c^3_{13}+c^4_{13} &\gamma_1-c^1_{13} & c^3_{13}+c^4_{13} &\gamma_1 & c^3_{13} & -1 \\
 -c^3_{13}-c^4_{13} & -1 & c^2_{13}+c^3_{13}+c^4_{13} & c^3_{13} &-\gamma_1 & \gamma_1-c^1_{13} \\
 c^3_{13} & -c^2_{13}-c^3_{13}-c^4_{13} &\gamma_1 & -1 & -\gamma_1+c^1_{13} &c^3_{13}+c^4_{13}\\
\end{smallmatrix}
\right)\\
 \gamma_1&= c^1_{13}+c^2_{13}+c^3_{13}+c^4_{13}+c^5_{13}\\
S_{00}&=0.55065, \qquad
 m_i=(1,1,1,1,1,1),\qquad
ind_p=5.
\end{align*}
\item \begin{align*}
    \bar{S}[13,2]&=
\left(
\begin{smallmatrix}
 1 & c^2_{13}+c^4_{13}+1 & c^1_{13}+c^3_{13}+c^5_{13} & c^1_{13} & c^2_{13}+1 & c^1_{13}+c^3_{13} \\
 c^2_{13}+c^4_{13}+1 & -1 & c^1_{13}+c^3_{13} & -c^2_{13}-1 & c^1_{13} & -c^1_{13}-c^3_{13}-c^5_{13} \\
 c^1_{13}+c^3_{13}+c^5_{13} & c^1_{13}+c^3_{13} & -c^2_{13}-1 & -1 & c^2_{13}+c^4_{13}+1 & -c^1_{13} \\
 c^1_{13} & -c^2_{13}-1 & -1 & c^1_{13}+c^3_{13} & c^1_{13}+c^3_{13}+c^5_{13} & -c^2_{13}-c^4_{13}-1 \\
 c^2_{13}+1 & c^1_{13} & c^2_{13}+c^4_{13}+1 & c^1_{13}+c^3_{13}+c^5_{13} & -c^1_{13}-c^3_{13} & -1 \\
 c^1_{13}+c^3_{13} & -c^1_{13}-c^3_{13}-c^5_{13} & -c^1_{13} & -c^2_{13}-c^4_{13}-1 & -1 & c^2_{13}+1 \\
\end{smallmatrix}
\right)\\
S_{00}&=0.25778, \qquad
 m_i=(1,1,1,1,1,1),\qquad
ind_p=2.
\end{align*}
\item \begin{align*}
    \bar{S}[13,3]&=
\left(
\begin{smallmatrix}
 1 & \gamma_1 & c^3_{13}+1 & -c^2_{13}-c^5_{13} & c^1_{13}+c^2_{13}+c^5_{13} & -c^5_{13} \\
 \gamma_1 & -1 & -c^5_{13} & -c^1_{13}-c^2_{13}-c^5_{13} & -c^2_{13}-c^5_{13} & -c^3_{13}-1 \\
 c^3_{13}+1 & -c^5_{13} & -c^2_{13}-c^5_{13} & 1 &-\gamma_1 & -c^1_{13}-c^2_{13}-c^5_{13} \\
 -c^2_{13}-c^5_{13} & -c^1_{13}-c^2_{13}-c^5_{13} & 1 & c^3_{13}+1 & c^5_{13} & \gamma_1 \\
 c^1_{13}+c^2_{13}+c^5_{13} & -c^2_{13}-c^5_{13} &-\gamma_1 & c^5_{13} & -c^3_{13}-1 & 1 \\
 -c^5_{13} & -c^3_{13}-1 & -c^1_{13}-c^2_{13}-c^5_{13} & \gamma_1 & 1 & c^2_{13}+c^5_{13} \\
\end{smallmatrix}
\right)\\
\gamma_1&=c^1_{13}+c^2_{13}+c^4_{13}+c^5_{13}\\
S_{00}&=0.36783, \qquad
 m_i=(1,1,1,1,1,1),\qquad
ind_p=3.
\end{align*}
\item \begin{align*}
    \bar{S}[15,1]&=
\left(
\begin{smallmatrix}
 1 & 2 & 2 & 2 & 2 & 2 \\
 1 & -c^1_5-1 & c^1_5 & 2 & c^1_5 & -c^1_5-1 \\
 2 & 2c^1_5 & c^1_5+1 & -2 & -2c^1_5-2 & -c^1_5 \\
 1 & 2 & -1 & -1 & 2 & -1 \\
 1 & c^1_5 & -c^1_5-1 & 2 & -c^1_5-1 & c^1_5 \\
 2 & -2c^1_5-2 & -c^1_5 & -2 & 2c^1_5 & c^1_5+1 \\
\end{smallmatrix}
\right)\\
S_{00}&=\frac{1}{\sqrt{15}}, \qquad
 m_i=(1,2,1,2,2,1),\qquad
ind_p=0.
\end{align*}

\item \begin{align*}
    \bar{S}[20,1]&=
\left(
\begin{smallmatrix}
 1 & 2 & 2 & 1 & 2 & 2 \\
 1 & c^1_{10} & c^1_{10}-1 & -1 & 1-c^1_{10} & -c^1_{10} \\
 1 & c^1_{10}-1 & -c^1_{10} & 1 & -c^1_{10} & c^1_{10}-1 \\
 1 & -2 & 2 & -1 & -2 & 2 \\
 1 & 1-c^1_{10} & -c^1_{10} & -1 & c^1_{10} & c^1_{10}-1 \\
 1 & -c^1_{10} & c^1_{10}-1 & 1 & c^1_{10}-1 & -c^1_{10} \\
\end{smallmatrix}
\right)\\
S_{00}&=\frac{1}{\sqrt{10}}, \qquad
 m_i=(1,2,2,1,2,2),\qquad
ind_p=0.
\end{align*}
\item \begin{align*}
    \bar{S}[21,1]&=
\left(
\begin{smallmatrix}
 1 & 2 c^1_{21}+2 & c^2_{21}+c^5_{21} & 2 & 2 c^2_{21}+2 c^5_{21} & c^1_{21}+1 \\
 c^1_{21}+1 & c^2_{21}+c^5_{21} & 1 & -c^1_{21}-1 & -1 & -c^2_{21}-c^5_{21} \\
 c^2_{21}+c^5_{21} & 2 & -c^1_{21}-1 & 2 c^2_{21}+2 c^5_{21} & -2 c^1_{21}-2 & 1 \\
 1 & -c^1_{21}-1 & c^2_{21}+c^5_{21} & -1 & -c^2_{21}-c^5_{21} & c^1_{21}+1 \\
 c^2_{21}+c^5_{21} & -1 & -c^1_{21}-1 & -c^2_{21}-c^5_{21} & c^1_{21}+1 & 1 \\
 c^1_{21}+1 & -2 c^2_{21}-2 c^5_{21} & 1 & 2 c^1_{21}+2 & 2 & -c^2_{21}-c^5_{21} \\
\end{smallmatrix}
\right)\\
S_{00}&=0.18936, \qquad
 m_i=(1,2,1,2,2,1),\qquad
ind_p=0.
\end{align*}
\item \begin{align*}
    \bar{S}[24,1]&=
\left(
\begin{smallmatrix}
 1 & 2 & 2 & 2 & 1 & 2 \\
 2 & 0 & 2 & 0 & -2 & -2 \\
 1 & 1 & -1 & -2 & 1 & -1 \\
 1 & 0 & -2 & 0 & -1 & 2 \\
 1 & -2 & 2 & -2 & 1 & 2 \\
 1 & -1 & -1 & 2 & 1 & -1 \\
\end{smallmatrix}
\right)\qquad
S_{00}=\frac{1}{\sqrt{12}}, \qquad
 m_i=(1,1,2,2,1,2),\qquad
ind_p=0,4.
\end{align*}
\item \begin{align*}
    \bar{S}[28,1]&=
\left(
\begin{smallmatrix}
 1 &c^1_{7} &c^1_{14}-c^1_{7} &c^1_{7} &c^1_{14}-c^1_{7} & 1 \\
c^1_{7} &c^1_{14}-c^1_{7} & 1 &c^1_{7}-c^2_{14} & -1 & -c^1_{7} \\
c^1_{14}-c^1_{7} & 1 & -c^1_{7} & -1 &c^1_{7} &c^1_{7}-c^2_{14} \\
c^1_{7} &c^1_{7}-c^2_{14} & -1 &c^1_{7}-c^2_{14} & -1 &c^1_{7} \\
c^1_{14}-c^1_{7} & -1 &c^1_{7} & -1 &c^1_{7} &c^1_{14}-c^1_{7} \\
 1 & -c^1_{7} &c^1_{7}-c^2_{14} &c^1_{7} &c^1_{14}-c^1_{7} & -1 \\
\end{smallmatrix}
\right)\  
S_{00}=0.41790\  , \ 
 m_i=(1,1,1,1,1,1),\quad
ind_p=2,4.
\end{align*}
\item \begin{align*}
    \bar{S}[28,2]&=
\left(
\begin{smallmatrix}
 1 &c^1_{14} & 1 &c^1_{7}+1 &c^1_{7}+1 &c^1_{14} \\
c^1_{14} &c^1_{7}+1 & -c^1_{14} & 1 & -1 & -c^1_{7}-1 \\
 1 & -c^2_{14} & -1 &c^1_{7}+1 & -c^1_{7}-1 &c^1_{14} \\
c^1_{7}+1 & 1 &c^1_{7}+1 & -c^1_{14} & -c^1_{14} & 1 \\
c^1_{7}+1 & -1 & -c^1_{7}-1 & -c^1_{14} &c^1_{14} & 1 \\
c^1_{14} & -c^1_{7}-1 &c^1_{14} & 1 & 1 & -c^1_{7}-1 \\
\end{smallmatrix}
\right)\ 
S_{00}=0.23192\ , \ 
 m_i=(1,1,1,1,1,1)\ ,\ 
ind_p=0,2.
\end{align*}
\item \begin{align*}
    \bar{S}[32,1]=
\left(
\begin{smallmatrix}
 1 & 2 c^1_{16} &  c^1_{8}+1 & c^1_{16}+ c^3_{16} & 1 &  c^1_{8}+1 \\
 c^1_{16} & 0 & c^1_{16} & 0 & - c^1_{16} & - c^1_{16} \\
  c^1_{8}+1 & 2 c^1_{16} & -1 & - c^1_{16}- c^3_{16} &  c^1_{8}+1 & -1 \\
 c^1_{16}+ c^3_{16} & 0 & - c^1_{16}- c^3_{16} & 0 & - c^1_{16}- c^3_{16} & c^1_{16}+ c^3_{16} \\
 1 & -2 c^1_{16} &  c^1_{8}+1 & - c^1_{16}- c^3_{16} & 1 &  c^1_{8}+1 \\
  c^1_{8}+1 & -2 c^1_{16} & -1 & c^1_{16}+ c^3_{16} &  c^1_{8}+1 & -1 \\
\end{smallmatrix}
\right)
\end{align*}
\[
S_{00}=\frac{c_{16}^3}4\ , \
 m_i=(1,2,1,1,1,1) \ ,\ 
ind_p=0,4.
\]
\item \begin{align*}
    \bar{S}[32,2]&=
\left(
\begin{smallmatrix}
  1 & c^1_8-1 & 1 & c^1_{16}-c^3_{16} & 2 c^3_{16} & c^1_8-1 \\
 c^1_8-1 & -1 & c^1_8-1 & c^1_{16}-c^3_{16} & -2 c^3_{16} & -1 \\
 1 & c^1_8-1 & 1 & c^3_{16}-c^1_{16} & -2 c^3_{16} & c^1_8-1 \\
 c^1_{16}-c^3_{16} & c^1_{16}-c^3_{16} & c^3_{16}-c^1_{16} & 0 & 0 & c^3_{16}-c^1_{16} \\
 c^3_{16} & -c^3_{16} & -c^3_{16} & 0 & 0 & c^3_{16} \\
 c^1_8-1 & -1 & c^1_8-1 & c^3_{16}-c^1_{16} & 2 c^3_{16} & -1 \\
\end{smallmatrix}
\right)\\
S_{00}&=\frac{c^1_{16}}{2}, \qquad
 m_i=(1,1,1,1,2,1),\qquad
ind_p=1,5.
\end{align*}
\item \begin{align*}
    \bar{S}[35,1]&=
\left(
\begin{smallmatrix}
 1 & c^1_7+1 & c^1_{35}+c^6_{35} & c^1_{5}+1 & c^1_7+c^2_{5}+1 & \gamma \\
 c^1_7+1 & -c^1_7-c^2_{5}-1 & c^1_{5}+1 & \gamma& 1 & -c^1_{35}-c^6_{35} \\
 c^1_{35}+c^6_{35} & c^1_{5}+1 & c^1_7+1 & -c^1_7-c^2_{5}-1 & -\gamma & -1 \\
 c^1_{5}+1 & \gamma & -c^1_7-c^2_{5}-1 & -1 & c^1_{35}+c^6_{35} & -c^1_7-1 \\
 c^1_7+c^2_{5}+1 & 1 & -c^1_{35}-\gamma & c^1_{35}+c^6_{35} & -c^1_7-1 & c^1_{5}+1 \\
 \gamma & -c^1_{35}-c^6_{35} & -1 & -c^1_7-1 & c^1_{5}+1 & c^1_7+c^2_{5}+1 \\
\end{smallmatrix}
\right)\\
\gamma&=c^1_{35}+c^4_{35}+c^6_{35}+c^{11}_{35}\ ,S_{00}=0.17243,\qquad
 m_i=(1,1,1,1,1,1),\qquad
ind_p=0.
\end{align*}

\item \begin{align*}
    \bar{S}[40,1]&=
\left(
\begin{smallmatrix}
1 & c^2_{10} & 2c^2_{10} & 1 & c^2_{10} & 2 \\
 c^2_{10} & -1 & 2 & c^2_{10} & -1 & -2c^2_{10}\\
 c^2_{10} & 1 & 0 & -c^2_{10} & -1 & 0 \\
 1 & c^2_{10} & -2c^2_{10} & 1 & c^2_{10} & -2 \\
 c^2_{10} & -1 & -2 & c^2_{10} & -1 & 2c^2_{10} \\
 1 & -c^2_{10} & 0 & -1 & c^2_{10} & 0 \\
\end{smallmatrix}
\right)\\
S_{00}&=0.26286,\qquad
 m_i=(1,1,2,1,1,2),\qquad
ind_p=0,3.
\end{align*}
\item \begin{align*}
    \bar{S}[48,1]&=
\left(
\begin{smallmatrix}
 1 & 2 & 2 c^2_8 & 1 & 2 & c^2_8 \\
 1 & -1 & c^2_8 & 1 & -1 & -c^2_8 \\
 c^2_8 & c^2_8 & 0 & -c^2_8 & -c^2_8 & 0 \\
 1 & 2 & -2 c^2_8 & 1 & 2 & -c^2_8 \\
 1 & -1 & -c^2_8 & 1 & -1 & c^2_8 \\
 c^2_8 & -2 c^2_8 & 0 & -c^2_8 & 2 c^2_8 & 0 \\
\end{smallmatrix}
\right)\\
S_{00}&=\frac{1}{2\sqrt{3}},\qquad
 m_i=(1,2,2,1,2,1),\qquad
ind_p=0,3.
\end{align*}
\item \begin{align*}
    \bar{S}[80,1]&=
\left(
\begin{smallmatrix}
 1 &\gamma & c_{10}^1 & \sqrt2 & 1 & c_{10}^1 \\
\gamma & 0 & \sqrt2 & 0 & -\gamma & -c_{10}^1 \\
 c_{10}^1 & \sqrt2 & -1 & -\gamma & c_{10}^1 & -1 \\
 \sqrt2 & 0 & -\gamma & 0 & -c_{10}^1 &\gamma \\
 1 & -\gamma & c_{10}^1 & -c_{10}^1 & 1 & c_{10}^1 \\
 c_{10}^1 & -c_{10}^1 & -1 &\gamma & c_{10}^1 & -1 \\
\end{smallmatrix}
\right)\\
 \gamma&=c^3_{40}+c^5_{40}-c^7_{40},\quad
S_{00}=0.26286,\quad
 m_i=(1,1,1,1,1,1),\quad
ind_p=0,4.
\end{align*}
\end{enumerate}
\bibliographystyle{jhep}
\bibliography{master} 

\end{document}